\documentclass{revtex4}
\usepackage{amssymb}
\usepackage{amsfonts}
\usepackage{amsmath}
\usepackage{graphicx}

\input{tcilatex}

\providecommand{\unit}[1]{\mathrm{#1}}

\begin{document}

\title{Advanced computational methods for simulating the classical hydrogen
atom in stochastic electrodynamics}
\author{Daniel C. Cole}
\affiliation{Boston University}

\begin{abstract}
Stochastic electrodynamics (SED) is a theory of nature seeking to derive key
aspects of quantum mechanics and quantum electrodynamics from a purely
classical physics perspective. While SED has been successful in many areas,
it has had mixed success when addressing the classical hydrogen atom bathed
in classical electromagnetic zero-point radiation. Because analytical
approaches are largely intractable, advanced numerical simulation methods
are required. \ To address this, we develop a pseudo-wave formulation that
integrates the solid angle in $k-$space utilizing symmetry, resulting in a
significantly reduced memory footprint and lower computational overhead for
generating the zero-point radiation spectrum. This optimization is paired
with a logarithmic frequency grid, a sliding window tracking the
instantaneous Keplerian frequency, and a Tukey smoothing function to
eliminate discrete force discontinuities. Integrated via an adaptive RK4
algorithm with step-doubling\ across highly dense grids of up to 100-million
bins, these techniques demonstrate the asymptotic convergence of ensemble
stability. We show that increasing grid resolution in the high-frequency
domain systematically mitigates artificial electron decay, yielding a
simulated probability density function that closely aligns with the quantum
mechanical ground state $P\left( r\right) $. \ Crucially, the simulations
exhibit stable, low-eccentricity orbits without any observed atomic
ionization events.
\end{abstract}

\maketitle

\section{Introduction}

This article pertains to a classical physics theory of nature usually
referred to as stochastic electrodynamics (SED). \ Most work in SED has been
carried out by applying analytical methods to linear and nonlinear
oscillator systems. \ However, such approaches have not been successful when
addressing physically realistic atomic systems. \ Consequently,
computational attempts have been made in Refs. \cite{ColeWilsonPRLett2003}, 
\cite{Cole2006-Vaxjo-Sweden}, \cite{Nieuwenhuizen2015-Nonrel}, \cite%
{Nieuwenhuizen2015-Relativistic}. \ These simulation studies have to do with
the stochastic equilibrium of the classical hydrogen atom (H), where a
classical electron is held bound to the nucleus by the Coulombic binding
force (proportional to $1/r^{2}$), and the atom is \textquotedblleft
bathed\textquotedblright\ in electromagnetic (EM) zero-point (ZP) radiation.
\ These two aspects of (1) a nonlinear binding force and the (2) stochastic
interaction between charged particle and radiation, make the simulation
problem difficult, particularly regarding problems of memory, speed, and
accuracy. This article addresses these points. \ The importance of resolving
these problems is no less than helping to learn whether SED may serve as a
fundamental basis for quantum mechanics (QM).

The goal of the simulation program that will be described here is to
determine whether a classical electron bound by the Coulombic force to the
classical nucleus/proton, plus bathed in EM ZP radiation, will result in a
classical H atom where the electron's orbit does not decay into the nucleus
(\textquotedblleft atomic collapse\textquotedblright ), nor does the
electron \textquotedblleft ionize\textquotedblright\ from the nucleus. \ The
program solves for the probability density $P\left( r\right) $ of the
electron being at a distance $r$ from the nucleus. \ If physically accurate,
then $P\left( r\right) $ should agree with the predicted result from QM:

\begin{equation}
P_{\text{QM}}\left( r\right) =\frac{4}{a_{\text{B}}^{3}}r^{2}\exp \left( -%
\frac{2r}{a_{\text{B}}}\right) \ \ ,  \label{Pqm(r)}
\end{equation}%
where $a_{\text{B}}$ is the Bohr radius.

The three main solutions described here for improving the simulation
capabilities in Refs. \cite{ColeWilsonPRLett2003}-\cite%
{Nieuwenhuizen2015-Relativistic} are: (1) A means using symmetry in
\textquotedblleft $\vec{k}-$ space\textquotedblright\ to greatly reduce the
number of random variables used in the computations, thereby greatly saving
considerable memory and improving speed issues, (2) a logarithmic gridding
method to improve speed and accuracy when dealing with radiation frequencies
that span five to six orders of magnitude, and (3) several means of
\textquotedblleft smoothing\textquotedblright\ out the extremely rapidly
varying stochastic force from the \textquotedblleft
zero-point\textquotedblright\ radiation, without sacrificing the essential
interaction between radiation and charged particles.\ \ These approaches
have not been used before in SED.

The first approach deals with the random variables that define the electric
and magnetic field values at points in space and time. \ Using the electric
dipole approximation (Sect. 3.0), it is shown that the random variables can
be combined using symmetry of angular integration, thereby enormously
cutting down on computational memory needs and improving speed. \ In terms
of memory the improvement is from $O\left( N^{3}\right) $ to $O\left(
N\right) $.

The second approach addresses the five to six orders of magnitude of the
frequencies required of the electric and magnetic fields. \ Instead of a
linear scaling of values, a logarithmic scaling is used (Sect. 4.1), which
cuts down on electromagnetic field iterations at each point in time.

The third method addresses near infinite variations in time and space of the
\textquotedblleft zero-point\textquotedblright\ radiation in two ways: The
first is to extract the main resonant part of the particle to radiation
interaction (Sect. 4.2). \ Although this has been done before \cite%
{ColeWilsonPRLett2003}, the method is improved here by using the log scaling
method just mentioned. \ The second part (Sects. 4.3-4.4) has to do with
making sure that possible discontinuities in the ZP Lorentz force
calculations are minimized.

The simulation studies in 2003 Ref. \cite{ColeWilsonPRLett2003} and 2006
Ref. \cite{Cole2006-Vaxjo-Sweden} deal with a \textquotedblleft 2D
version\textquotedblright\ of the classical H atom. The present 2026 code
agrees with the 2003/2006 work, but has now been developed for the full 3D
situation. \  All three of these works differ considerably versus the 2015
3D H simulation work of Refs. \cite{Nieuwenhuizen2015-Nonrel} and \cite%
{Nieuwenhuizen2015-Relativistic}. \ References \cite%
{Nieuwenhuizen2015-Nonrel} and \cite{Nieuwenhuizen2015-Relativistic} predict
rapidly changing eccentric orbits peaking as high as 0.8, and more
importantly, that ionization occurs. \ This did not occur in the 2003/2006
codes, nor in the present code. \ Indeed, the eccentricity in these latter
works remains very low, close to zero, which is that of nearly circular
orbits. \ Likely the difference is the third point mentioned above of
dealing with the seemingly discontinuous trajectories of the electron. \ No
ionizations are predicted for the present code, nor for the 2003/2006 work.
\ This will be described in more detail here in Sects. 4.2-4.4.

\ The \textquotedblleft smoothing\textquotedblright\ of the Lorentz force is
done here using a Tukey cosine function (Sect. 4.3). \ Together, all of
these elements in Sects. 4.2-4.4.4 eliminate discrete step-changes in the
simulated zero-point radiation, preventing the catastrophic numerical shocks
that otherwise cause adaptive differential equation solvers to fail or
predict spurious ionization-like effects. \ To a large extent these results
are greatly aided with the first and second methods just mentioned to
provide the necessary speed and memory improvements.

In addition, further techniques that have helped improve the simulation
approach are discussed here, including evaluation of the ZP Lorentz force at
all integration points and a sliding window tracking the instantaneous
Keplerian frequency. \ The trajectory integration of the electron is done in
our C++ code via an adaptive Runge-Kutta fourth order (RK4) algorithm with
\textquotedblleft step-doubling\textquotedblright\ (also known as Richardson
Extrapolation) \cite{Press2007NumericalRecipes} across highly dense grids of
up to 100-million bins. \ These methods all combine to make the simulation
code considerably faster and more accurate in its predictions.

The outline of this article is as follows: Sect. 2.1 presents a brief
overview of SED for those not familiar with the theory, while Sect. 2.2
describes the typical starting point for mathematical analysis in SED. \
Section 2.3 discusses why the two main mathematical methods in Sects. 3.0
and 4.0 are important ones for aiding simulation computations in SED for the
hydrogen atom and future atomic systems. \ Section 3.0 turns to the first of
two main algorithm improvements, namely, a means using symmetry in
\textquotedblleft $k-$space\textquotedblright\ to greatly reduce the number
of random variables used in the computations. Section 4.0 turns to a scaling
means for working with frequency variables that span orders of magnitude in
number, plus goes over a number of other computational improvements. \
Section 5.0 describes work carried out during the course of the development
of the simulation software. \ This analysis helped confirm and direct the
code/simulation creation to make sure the predicted physics and algorithms
were working properly. \ Section 6.0 describes the computer code in more
detail and shows examples of what can be obtained from the simulator.\ \
Section 7.0 ends with some concluding remarks. \ Ultimately, this article
provides a comprehensive blueprint of the simulator's architecture, enabling
researchers to leverage this publicly available software for advanced
physical analysis in SED \cite{Cole-Alpha-2026}.

\section{Overview of SED}

\subsection{Background}

We begin here with a brief overview of SED, including the variety of
problems that have been successfully tackled, but also the difficulties
encountered when moving onto more physically realistic atomic systems. \ The
focus here will be on significant algorithm improvements that have made
computer simulations of classical atomic hydrogen more feasible.

SED is a \textquotedblleft classical\textquotedblright\ theory of nature,
not in the sense that might initially be perceived, but rather in that it
only involves the physics of the special relativistic version of Newton's
second law of nature as applied to point charged particles (so the
Lorentz-Dirac equation \cite{Teitelboim1970},\cite{teitelboim}), and of
Maxwell's equations. \ Hence SED consists of \textquotedblleft
classical\textquotedblright\ electrodynamics up to the approximate time of
QM. \ It also assumes that classical EM radiation can exist at temperature $%
T=0$ and in fact is essential to having equilibrium (\textit{i.e.},
stochastic equilibrium) between classical charged particles and classical
EM\ radiation. \ This radiation at $T=0$ is referred to as classical
electromagnetic ZP radiation. \ These few points have resulted in
predictions in physical phenomena that are normally thought to only be
addressable by QM and quantum electrodynamics (QED). \ Indeed, some
physicists, including this author, have felt that many QM and QED aspects
may be viewed as derivable from SED in the sense that thermodynamics is
derivable from statistical mechanics.

Several reviews of SED exist such as Refs. \cite{delaPena1996} (SED up until
1996), \cite{Boyer1975review1}, \cite{Cole2006-Vaxjo-Sweden} (first half of
article), and \cite{BoyerAtoms2019}. \ QM\ and QED\ phenomena that SED\ has
been successful in addressing are: (1) Ref. \cite{Boyer1975review2} compares
properties of the simple harmonic oscillator (SHO) within SED, QM, and QED;
(2)\ van der Waals forces have been calculated to all distances and all
temperatures \cite{Boyer1973}, \cite{boyer--vanderWaals}, \cite{Cole1986};
(3) Casimir forces \cite{delaPena1996}, \cite{BoyerCasimir2003}; (4)
diamagnetism \cite{Boyer1980diam}; and (5) the thermal effects of
acceleration \cite{BoyerThermalEff1980}, \cite{BoyerThermalEff1984}, \cite%
{BoyerThermalEffDia1984}, \cite{Cole1985}, \cite{Cole1986}, \cite{Cole1987}.
\ Yet more examples could certainly be cited.

Prior to the early 2000's the SED research largely involved applying various
analytic methods to linear and nonlinear oscillator systems. \ However these
analytic methods were not feasible for the nonlinear Coulombic binding force
in atoms, combined with stochastic electromagnetic (EM) radiation bathing
the system. \ 

In 2003 a computer simulation method was developed to successfully move past
this barrier \cite{ColeWilsonPRLett2003},\cite{Cole2006-Vaxjo-Sweden}. Very
reasonable results were obtained as compared with quantum mechanics (QM). \
However, many approximations were made to enable feasible computations. \ In
particular, one approximation involved making the atomic simulation be a
\textquotedblleft two dimensional (2D)\textquotedblright\ model. \
Fortunately in 2015 sufficient computer resources were garnered with
improved algorithms to attack the problem in 3D \cite%
{Nieuwenhuizen2015-Nonrel},\cite{Nieuwenhuizen2015-Relativistic}. \
Computational issues have still persisted however. \ This article describes
several key improvements in Sects. 3.0 and 4.0 to overcome these problems.

\subsection{Mathematical start to SED}

The computational framework of SED can be summed up via the following
mathematical physics based descriptions of: (1) the classical
electromagnetic fluctuating fields of radiation in space, and (2) the
equations governing a system of classical charged point particles within
this radiation. Regarding (2), the Lorentz-Dirac equation \cite%
{Teitelboim1970}, \cite{teitelboim}, \cite{Plass1961}, \cite{Rohrlich} in
special relativity is the key equation for the motion of these classical
charged point particles, largely intended to represent electrons. \ Going
beyond point particles is certainly of interest in SED. \ However, there has
been less research pursued here as the charged point particles research has
been difficult enough, and point particles are a natural description for
electrons and a fairly good model for nuclei as well, at least at the atomic
level.

The radiation fields analyzed here will be described by filling a large
region of space, such that a classical model of an atomic system of nuclei
and electrons in space are considerably smaller than the dimensions of this
volume, and that the wavelength of the resonant frequency of the fluctuating
electrons are similarly much smaller. \ At the end of our calculations here,
the volume is usually extended to infinity. \ For simplicity of
calculations, the volume of radiation is usually taken to be a rectangular
parallelepiped region that the radiation fields are confined within. \
However, other shapes have been considered \cite{Cole2000ConductCav}. \ The
rectangular parallelepiped considered here will have dimensions $L_{x}$, $%
L_{y}$, and $L_{z}$, along the $x$, $y$, and $z$ axes, respectively.

The radiation fields will be represented as an infinite sum of plane waves,
with periodic boundary conditions imposed. \ \ Thus we can then use Fourier
series to describe the fields. \ Here we will use the following sum of plane
waves for the electric $\vec{E}_{\text{rad}}\left( \vec{x},t\right) $ and
magnetic $\vec{B}_{\text{rad}}\left( \vec{x},t\right) $ radiation fields in
this large rectangular parallelepiped volume:

\begin{equation}
\vec{E}_{\text{rad}}\left( \vec{x},t\right) =\frac{1}{\left(
L_{x}L_{y}L_{z}\right) ^{1/2}}\sum\limits_{n_{x},n_{y},n_{z}=-\infty
}^{\infty }\sum\limits_{\lambda =1,2}\hat{\varepsilon}_{\vec{n},\lambda }%
\left[ 
\begin{array}{c}
A_{\vec{n},\lambda }\cos \left( \vec{k}_{\vec{n}}\cdot \vec{x}\mathbf{-}%
\omega _{\vec{n}}t\right) \\ 
+B_{\vec{n},\lambda }\sin \left( \vec{k}_{\vec{n}}\cdot \vec{x}\mathbf{-}%
\omega _{\vec{n}}t\right)%
\end{array}%
\right] \ \ ,  \label{E-in}
\end{equation}%
\begin{equation}
\vec{B}_{\text{rad}}\left( \vec{x},t\right) =\frac{1}{\left(
L_{x}L_{y}L_{z}\right) ^{1/2}}\sum\limits_{n_{x},n_{y},n_{z}=-\infty
}^{\infty }\sum\limits_{\lambda =1,2}\left( \vec{k}_{\vec{n}}\mathbf{\times }%
\hat{\varepsilon}_{\vec{n},\lambda }\right) \left[ 
\begin{array}{c}
A_{\vec{n},\lambda }\cos \left( \vec{k}_{\vec{n}}\cdot \vec{x}\mathbf{-}%
\omega _{\vec{n}}t\right) \\ 
+B_{\vec{n},\lambda }\sin \left( \vec{k}_{\vec{n}}\cdot \vec{x}\mathbf{-}%
\omega _{\vec{n}}t\right)%
\end{array}%
\right] \ \ ,  \label{B-in}
\end{equation}%
where%
\begin{equation}
\vec{k}_{\vec{n}}=\frac{2\pi n_{x}}{L_{x}}\hat{x}\mathbf{+}\frac{2\pi n_{y}}{%
L_{y}}\hat{y}\mathbf{+}\frac{2\pi n_{z}}{L_{z}}\hat{z}\mathbf{\ \ ,}
\label{kn}
\end{equation}%
$n_{x}$, $n_{y}$, and $n_{z}$ are integers and $\vec{n}=n_{x}\hat{x}\ 
\mathbf{+\ }n_{y}\hat{y}\ \mathbf{+\ }n_{z}\hat{z}$ \ To satisfy all four of
Maxwell's classical electrodynamic equations for fields in free space, then $%
\omega _{\vec{n}}=c\left\vert \vec{k}_{\vec{n}}\right\vert $, $c$ is the
speed of light, $\vec{k}_{\vec{n}}\cdot \hat{\varepsilon}_{\vec{n},\lambda }=%
\vec{k}_{\vec{n}}\cdot \hat{\varepsilon}_{\vec{n},\lambda ^{\prime }}=0$,
and $\hat{\varepsilon}_{\vec{n},\lambda }\cdot \hat{\varepsilon}_{\vec{n}%
,\lambda ^{\prime }}=0$ for $\lambda \neq \lambda ^{\prime }$, where $%
\lambda $ and $\lambda ^{\prime }$ indicate the linear polarization
direction. \ Here, $\lambda $ and $\lambda ^{\prime }$ are indices that take
on only two values, so each might be represented by the values 1 or 2, for
example. \ Moreover, $\hat{k}_{\vec{n}}=\vec{k}_{\vec{n}}/\left\vert \vec{k}%
_{\vec{n}}\right\vert $; similarly, all other vectors with a
\textquotedblleft caret\textquotedblright\ are meant to be unit vectors.

It should be noted that one can show, using the free space Maxwell's
equations, that Eqs. (\ref{E-in}) and (\ref{B-in}) satisfy the wave
equations of $\nabla ^{2}\vec{E}\left( \vec{x},t\right) =\frac{1}{c^{2}}%
\frac{\partial ^{2}}{\partial t^{2}}\vec{E}\left( \vec{x},t\right) $ and $%
\nabla ^{2}\vec{B}\left( \vec{x},t\right) =\frac{1}{c^{2}}\frac{\partial ^{2}%
}{\partial t^{2}}\vec{B}\left( \vec{x},t\right) $. \ Moreover, the presence
of $\hat{\varepsilon}_{\vec{n},\lambda }$ and $\left( \vec{k}_{\vec{n}}%
\mathbf{\times }\hat{\varepsilon}_{\vec{n},\lambda }\right) $ in Eqs. (\ref%
{E-in}) and (\ref{B-in}), plus the cited relationships of $\vec{k}_{\vec{n}%
}\cdot \hat{\varepsilon}_{\vec{n},\lambda }=\vec{k}_{\vec{n}}\cdot \hat{%
\varepsilon}_{\vec{n},\lambda ^{\prime }}=0$, and $\hat{\varepsilon}_{\vec{n}%
,\lambda }\cdot \hat{\varepsilon}_{\vec{n},\lambda ^{\prime }}=0$ for $%
\lambda \neq \lambda ^{\prime }$, enable all four free space Maxwell's
equations to be satisfied.

The stochastic nature of the radiation fields in Eqs. (\ref{E-in}) and (\ref%
{B-in}), as prescribed in SED, arises from the probability distribution of
the Fourier coefficients $A_{1},\cdots ,A_{N},B_{1},\cdots ,B_{N}$ over a
large \textquotedblleft ensemble\textquotedblright\ of equally sized space
regions, each of which have the same dimensions of $L_{x},L_{y},L_{z}$. \
The radiation in the rectangular parallelepiped volumes in the ensemble are
each characterized by different Fourier coefficients, resulting in a net
probability distribution for each Fourier coefficient. \ It is in this way
that the word \textquotedblleft stochastic\textquotedblright\ enters into
the name for SED. \ The radiation acts on charged particles, making the
charged particles move differently in each member of the ensemble of
radiation. \ The result is that the particles in the ensemble display
stochastic properties as well. \ Each set of electrons and nuclei, plus the
radiation fields, represents one member of the ensemble of moving electrons,
nuclei, and fields.

In SED, the Fourier coefficients $A_{1},\cdots ,A_{N},B_{1,}\cdots ,B_{N}$
are assumed to be independent random variables with a Gaussian probability
density distribution:%
\begin{equation}
P_{\text{F}}\left( A_{p}\right) =\frac{1}{\sqrt{2\pi \left( \sigma
_{p}\right) ^{2}}}\exp \left[ -\frac{A_{p}^{2}}{2\left( \sigma _{p}\right)
^{2}}\right] \text{ \ ,}  \label{P(A-i)}
\end{equation}%
where the \textquotedblleft $p$\textquotedblright\ index of $p=1,2,3,...$ is
a shorthand notation for all the indices of $\vec{k}_{\vec{n}},\lambda $ in,
for example, $A_{\vec{n},\lambda }$. \ The same probability density
distribution in Eq. (\ref{P(A-i)}) also holds for the Fourier coefficient $%
B_{p}$. \ The label \textquotedblleft F\textquotedblright\ is added to
specify that this probability density function $P_{\text{F}}$ refers to the
Fourier coefficients. \ Moreover, $\sigma _{p}$ depends on $\left\vert \vec{k%
}_{\vec{n}}\right\vert =$ $\omega _{\vec{n}}/c$ and therefore on $\left\vert 
\vec{n}\right\vert $ as in Eq. (\ref{kn}). The functional form of $\sigma
_{p}$ at $T=0$ is a cornerstone for SED and expressed by 
\begin{equation}
\sigma ^{2}\left( \omega _{\vec{n}},T=0\right) =2\pi \hbar \omega _{\vec{n}%
}\ \ ,  \label{sig-i}
\end{equation}%
where $\sigma _{p}^{2}$ in Eq. (\ref{P(A-i)}) means this quantity in Eq. (%
\ref{sig-i}), and $\hbar $ is Planck's reduced constant.\ It is this simple
linear function of $\omega $ at $T=0$ that provides the Lorentz invariance
property deduced by Marshall \cite{Marshall1965a} and Boyer \cite%
{Boyer1969Black1}, and much later in Refs. \cite{Cole19901st}, \cite%
{Cole-3-charges-2025}, \cite{Cole-dipole-2026} that Eq. (\ref{sig-i})
results in no heat flowing during isothermal, reversible processes . \
Imposing either of these properties onto Eqs. (\ref{E-in}) and (\ref{B-in})
enables the linear function of $\omega $ in Eq. (\ref{sig-i}) to be deduced,
where the $2\pi \hbar $ factor arises by comparing with measurements of
various physical phenomena.

When $T>0$, the main spectrum of interest in SED is the \textquotedblleft
zero-point plus Planckian (ZPP)\textquotedblright\ radiation spectrum, 
\begin{equation}
\sigma ^{2}\left( \omega _{\vec{n}},T\right) =2\pi \hbar \omega _{\vec{n}}+%
\frac{4\pi \hbar \omega _{\vec{n}}}{\exp \left( \frac{\hbar \omega _{\vec{n}}%
}{k_{\text{B}}T}\right) -1}=2\pi \hbar \omega _{\vec{n}}\coth \left( \frac{%
\hbar \omega _{\vec{n}}}{2k_{\text{B}}T}\right) \ \ ,  \label{ZPP}
\end{equation}%
where $k_{\text{B}}$ is Boltzmann's constant.\ \ When $T=0$, for the ZP
case, this reduces to Eq.(\ref{sig-i}), $\sigma _{p}^{2}=2\pi \hbar \omega _{%
\vec{n}}$. \ For Rayleigh-Jeans (RJ) radiation, $\sigma _{p}^{2}=4\pi k_{%
\text{B}}T$, which is what the ZPP expression in Eq. (\ref{ZPP}) becomes
when $k_{\text{B}}T\gg \hbar \omega _{\vec{n}}\ $.

Important properties of the $A$ and $B$ discrete variables are that%
\begin{equation}
\left\langle A_{\vec{n},\lambda }\right\rangle =\left\langle B_{\vec{n}%
},\lambda \right\rangle \ \ \text{\ for all }\vec{n}\ \ .  \label{<A>}
\end{equation}%
\begin{equation}
\left\langle A_{\vec{n},\lambda }B_{\vec{n}^{\prime },\lambda ^{\prime
}}\right\rangle =0\text{ for any values of }\vec{n}\text{, }\vec{n}^{\prime }%
\text{, }\lambda \text{, and }\lambda ^{\prime }.  \label{<AB>}
\end{equation}%
\begin{equation}
\left\langle A_{\vec{n},\lambda }A_{\vec{n}^{\prime },\lambda ^{\prime
}}\right\rangle =\delta _{\lambda \lambda ^{\prime }}\delta _{\vec{n}\vec{n}%
^{\prime }}2\pi \hbar \omega _{\vec{n}}\ \ ,  \label{<AA'>}
\end{equation}%
The $\left\langle \cdots \right\rangle $ brackets mean to take an ensemble
average over all the members (radiation ZP field plus electron trajectory)
of the ensemble. \ For each member of this ensemble, the $A^{\prime }s$ and $%
B^{\prime }s$ values are fixed. \ However, they are \textquotedblleft
randomly\textquotedblright\ different from one ensemble member to another,
but with Eq. (\ref{P(A-i)}) describing their distribution of values in the
net ensemble. \ Much of the above three equations is already embedded in the
statement accompanying Eq. (\ref{P(A-i)}) that, \textquotedblleft In SED,
the Fourier coefficients $A_{1},\cdots ,A_{N},B_{1,}\cdots ,B_{N}$ are
assumed to be \textbf{independent} random variables ...\ \
.\textquotedblright

The probability density of finding the fluctuating charged particle at
position $\vec{x}$ in the steady state condition is \cite%
{ColePhysicsTwo-Methods-2023}%
\begin{align}
P_{\text{3x}}\left( \vec{x}\right) & =\int\limits_{-\infty }^{\infty
}dA_{1}\cdots \int\limits_{-\infty }^{\infty }dA_{N}\cdots
\int\limits_{-\infty }^{\infty }dB_{1}\cdots \int\limits_{-\infty }^{\infty
}dB_{N}  \notag \\
& \text{\ \ \ \ \ \ }\times P_{\text{F}}\left( A_{1}\right) \cdots P_{\text{F%
}}\left( A_{N}\right) P_{\text{F}}\left( B_{1}\right) \cdots P_{\text{F}%
}\left( B_{N}\right) \delta ^{3}\left[ \vec{x}-\vec{x}_{A_{1},\cdots
,B_{N}}\left( t\right) \right] \ \ ,  \label{P(x)}
\end{align}%
where $\delta ^{3}\left[ \vec{x}-\vec{x}_{A_{1},\cdots ,B_{N}}\left(
t\right) \right] $ is the Dirac delta function in 3D space and the
multiplication of all the $P_{\text{F}}\left( A_{i}\right) $ and $P_{\text{F}%
}\left( B_{i}\right) $ functions is due to $A_{1},\cdots ,A_{N},B_{1,}\cdots
,B_{N}$ being independent random variables. \ At the end of the calculations
with Eq. (\ref{P(A-i)}), $N\rightarrow \infty $ \ is imposed. \ Finally, $%
\vec{x}_{A_{1},\cdots ,B_{N}\cdots }\left( t\right) $ in Eq. (\ref{P(x)}),
gives the trajectory of the particle and is a function of $A_{1},\cdots
,A_{N},B_{1},\cdots ,B_{N}$. \ All \textquotedblleft $A$\textquotedblright\
and \textquotedblleft $B$\textquotedblright\ coefficients are real
quantities and are integrated from $-\infty $ to $+\infty $ in Eq. (\ref%
{P(x)}). \ Their values control the steady state solution for the particle's
trajectory of $\vec{x}_{A_{1},\cdots ,B_{N}}\left( t\right) $. \ The
probability density of these fields, $P_{\text{F}}\left( A_{1}\right) ...P_{%
\text{F}}\left( A_{N}\right) P_{\text{F}}\left( B_{1}\right) ...P_{\text{F}%
}\left( B_{N}\right) $, dictates their contribution regarding $\delta ^{3}%
\left[ \vec{x}-\vec{x}_{A_{1},\cdots ,B_{N}}\left( t\right) \right] $ by
selecting\ the contribution to the final particle probability density $P_{%
\text{3x}}\left( \vec{x}\right) $. \ More explanation on how to carry out
such stochastic calculations involving the $A_{\vec{n},\lambda }$ and $B_{%
\vec{n},\lambda }$ coefficients is discussed in Refs. \cite%
{ColePhysicsTwo-Methods-2023},\cite{ColeAtoms2019},\cite%
{Cole-Frontier-SED-2021}.

\subsection{Why the computational methods aid in atomic solutions in SED}

As discussed in Sect. I, computational methods now exist to solve problems
of classical atoms that are represented not just by linear and nonlinear
oscillators, but by more realistic atomic systems. \ These methods have so
far only been applied to the simplest atom, namely, the classical hydrogen
(H) system \cite{ColeWilsonPRLett2003},\cite{Cole2006-Vaxjo-Sweden},\cite%
{Nieuwenhuizen2015-Nonrel},\cite{Nieuwenhuizen2015-Relativistic}. \ Still,
these methods look extendable to more complicated systems.

For just a single H atom, the frequencies of the electron orbiting the
nucleus can range from the low frequency orbits of an electron in a Rydberg
atom with large $r$, to the high frequencies of an electron in a close orbit
to the nucleus. \ For the latter, the orbits and speeds would need to be
described by special relativity (\textit{i.e.}, the Lorentz-Dirac equation).
\ The fluctuating motion of the orbiting electron dictates which frequencies
of the EM radiation in Eqs. (\ref{E-in}) and (\ref{B-in}) are the most
important ones to include. \ The frequency range of electromagnetic
radiation emitted from an excited state of a hydrogen atom spans from about $%
2\times 10^{9}\unit{Hz}$ in the radio wave spectrum to about $3\times 10^{15}%
\unit{Hz}$ in the UV spectrum, so 6 orders of magnitude. \ Sections 3.0 and
4.1 in this article address this point. \ Saving the values of $A_{1},\cdots
,A_{N},B_{1,}\cdots ,B_{N}$ over the entire frequency range of a simulation,
introduces a severe problem that Sect. 3.0 deals with. \ Section 4.1
addresses the difficulty of summing over the related needed quantities at
each time step, creating in particular a bottleneck at the high frequency
end. \ Here the results of Sect. 3.0 are used to arrive at a better means in
Sect. 4.1 for \textquotedblleft indexing\textquotedblright\ the scale of the 
$10^{6}$ range of variables. \ These computational methods reduce the memory
storage and the amount of time retrieving and summing over the different
parts of this information at each time step by many orders of magnitude.

\section{Wave-Vector Shell Integration}

\subsection{General idea}

The core algorithmic optimization in this model leverages spherical symmetry
to integrate the 3D $k-$space field modes over all solid angles for a given
frequency interval $\omega $ to $\omega +d\omega $. This procedure
drastically reduces the number of random Gaussian variables that must be
stored, retrieved, and summed at each integration time step, from scaling
with $\omega ^{2}$ to $\omega ^{0}$ (\textit{i.e.}, where $\omega ^{0}$
means no scaling with $\omega $, but just dependent as usual on $\Delta
\omega $).

The $A$ and $B$ variables in Eqs. (\ref{E-in}) and (\ref{B-in}) are
generated as random numbers with a Gaussian probability density. The random
variables $A$ multiplies the $\cos \left( \omega t\right) $ term and the $B$
random variable multiplies the $\sin \left( \omega t\right) $ term. \ \ As
described here, by taking into account the vector nature of the polarization
vectors, adding them all up and getting the net mean and variance, using
symmetry, enables the net to be described by only 12 random numbers for each
frequency interval of $\omega \rightarrow \omega +\Delta \omega $.

Moreover, in the large $L$ limit, where $L_{x}=L_{y}=L_{z}=L$ in Eqs. (\ref%
{E-in}) and (\ref{B-in}), $L$ cancels out of the analysis, where $L$ is the
length of a side of a large cube in 3D space. \ This makes perfect physical
sense since the physical behavior of the electron should not depend on the
periodically imposed value of $L$, which is just a mathematical means of
dealing with the system. \ Moreover, although $L_{x},L_{y},L_{z}$ need not
be restricted in our analysis to the single value $L$, analyzing a single
classical hydrogen in free space makes this assumption of $%
L_{x}=L_{y}=L_{z}=L$ also physically reasonable. \ (For examples in SED
where this restriction is not made, see Refs. \cite{ColeWilsonPRLett2003}
and \cite{Cole2000ConductCav}.)

$L$ drops out of the analysis as $L\rightarrow \infty $ due to two factors.
\ The first is that the number of infinitesimal volume cells of size $\left( 
\frac{2\pi }{L}\right) ^{3}$ within a large volume of a shell of $k$ space
is equal to 
\begin{equation*}
\frac{4\pi k^{2}dk}{\left( \frac{2\pi }{L}\right) ^{3}}\,.
\end{equation*}%
\ 

This brings up the second point: Each small volume $\left( \frac{2\pi }{L}%
\right) ^{3}$ that encloses a point of the $k$ vector in $k$ space, has an
electric field amplitude $A$ and $B$ from Eqs. (\ref{E-in}) and (\ref{B-in}%
). \ Each amplitude is weighted by $1/\sqrt{V}=1/L^{3/2}$, so the variance
involving the square of these random Gaussian amplitudes is weighted by $%
1/L^{3}$ (the sum of the means is zero). The number of $k$ points is
weighted by $L^{3}$ from above, so the net dependence on $L$ cancels out. \
This idea makes use of the fact that the sum of Gaussian random variables
gives a Gaussian random number with the mean given by the sum of the means
of the individual random numbers, and the net variance being the sum of the
individual variances.

\subsection{The statistical properties of the ZP electric field acting on
the classical electron}

The nonrelativistic ODE equation for a classical electron in a classical H
atom is given by nucleus to the electron at time $t\,$, 
\begin{equation}
m\frac{d^{2}}{dt^{2}}\vec{z}\left( t\right) =-\frac{e^{2}}{\left\vert \vec{z}%
\right\vert ^{3}}\hat{z}+\vec{F}_{\text{rr}}\left[ \vec{z}\left( t\right) ,t%
\right] -e\vec{E}_{\text{ZP}}\left[ \vec{z}\left( t\right) ,t\right] -\frac{e%
}{c}\frac{d\vec{z}}{dt}\times \vec{B}_{\text{ZP}}\left[ \vec{z}\left(
t\right) ,t\right] \ \ ,  \label{Eq Motion 1}
\end{equation}%
where $\vec{z}\left( t\right) $ is the vector distance from the nucleus, $m$
and $-e$ are the mass and charge of the classical electron, and 
\begin{equation}
\vec{F}_{\text{rr}}\left[ \vec{z}\left( t\right) ,t\right] =m\tau \frac{d^{3}%
\vec{z}}{dt^{3}}\ \ ,  \label{Frr}
\end{equation}%
is the nonrelativistic radiation reaction (see for example Ref. \cite%
{Jackson2}), where $\tau =\frac{2e^{2}}{3mc^{3}}$. \ In the case of H, the
common approximation to Eq. (\ref{Frr}) is:%
\begin{equation}
m\tau \frac{d^{3}\vec{z}}{dt^{3}}=m\tau \frac{d}{dt}\left( \frac{d^{2}\vec{z}%
}{dt^{2}}\right) \approx m\tau \frac{d}{dt}\left( -\frac{e^{2}}{m\left\vert 
\vec{z}\right\vert ^{3}}\hat{z}\right) \ \ \text{, }  \label{Frr Coul 2}
\end{equation}%
due to the strength of the binding force $-\frac{e^{2}}{\left\vert \vec{z}%
\right\vert ^{3}}\hat{z}$, or assuming that 
\begin{equation}
m\frac{d^{2}\vec{z}}{dt^{2}}\approx -\frac{e^{2}}{\left\vert \vec{z}%
\right\vert ^{2}}\hat{z}\ \ ,  \label{Frr Coul 3}
\end{equation}%
ignoring the contributions of $-e\vec{E}_{\text{ZP}}\left[ \vec{z}\left(
t\right) ,t\right] -\frac{e}{c}\frac{d\vec{z}}{dt}\times \vec{B}_{\text{ZP}}%
\left[ \vec{z}\left( t\right) ,t\right] \ \ .$ \ More will be said about
this in Sec. 5.4.

One can show that Eq. (\ref{Frr Coul 2}) results in 
\begin{equation}
\vec{F}_{\text{rr}}\left[ \vec{z}\left( t\right) ,t\right] =m\tau \frac{d^{3}%
\vec{z}}{dt^{3}}\approx \tau e^{2}\left[ -\frac{\frac{d\vec{z}}{dt}}{%
\left\vert \vec{z}\right\vert ^{3}}+\frac{3\vec{z}\mathbf{(}\vec{z}\mathbf{%
\cdot }\frac{d\vec{z}}{dt})}{\left\vert \vec{z}\right\vert ^{5}}\right] \ \ ,
\label{Frr Coul 4}
\end{equation}%
for the classical H case.

The probability/statistical aspects of the electron's trajectory originate
from $\vec{E}_{\text{ZP}}$ and $\vec{B}_{\text{ZP}}$ and are passed onto the
electron via Eq. (\ref{Eq Motion 1}). \ If we were treating the electron's
trajectory via the full relativistic equation of motion (Lorentz-Dirac
equation), the same basic idea would be true, meaning that the ZP fields's
probability character will be transferred to the electron's trajectory in
the same way. \ In a full analysis, the motion of the electron, with its
accelerations and decelerations radiating energy, works in concert with the
ZP\ field to influence its properties, particularly in a closed container. \
However, here we do not have a \textquotedblleft closed container; the
classical H atom is in empty space, but \textquotedblleft
bathed\textquotedblright\ in ZP field that then dictates the stochastic
properties of the electron's trajectory.

\ We will apply the \textquotedblleft dipole
approximation\textquotedblright\ to help us obtain approximate solutions of
Eq. (\ref{Eq Motion 1}). \ This step is a physical approximation that will
help us create computer code to numerically obtain such solutions of Eq. (%
\ref{Eq Motion 1}) \cite{Plass1961}. \ It consists of the following change
of Eq. (\ref{Eq Motion 1}):%
\begin{equation}
m\frac{d^{2}}{dt^{2}}\vec{z}\left( t\right) =-\frac{e^{2}}{\left\vert \vec{z}%
\right\vert ^{2}}\hat{z}+\vec{F}_{\text{rr}}-e\vec{E}_{\text{ZP}}\left(
0,t\right) -\frac{e}{c}\frac{d\vec{z}}{dt}\times \vec{B}_{\text{ZP}}\left(
0,t\right) \ \ .  \label{Eq Motion 2}
\end{equation}

In Eq. (\ref{Eq Motion 2}), and from Eqs. (\ref{E-in}) and (\ref{B-in}), the
dipole approximation means that 
\begin{equation}
\vec{E}_{\text{ZP}}\left( \vec{x},t\right) \approx \vec{E}_{\text{ZP}}\left(
0,t\right) =\frac{1}{\left( L\right) ^{3/2}}\sum\limits_{n_{x},n_{y},n_{z}=-%
\infty }^{\infty }\ \sum\limits_{\lambda =1,2}\hat{\varepsilon}_{\vec{n}%
,\lambda }\left[ A_{\vec{n},\lambda }\cos \left( \omega _{\vec{n}}t\right)
-B_{_{\vec{n},\lambda }}\sin \left( \omega _{\vec{n}}t\right) \right] \ \ ,
\label{E-in 2}
\end{equation}%
\begin{equation}
\vec{B}_{\text{ZP}}\left( \vec{x},t\right) \approx \vec{B}_{\text{ZP}}\left(
0,t\right) =\frac{1}{\left( L\right) ^{3/2}}\sum\limits_{n_{x},n_{y},n_{z}=-%
\infty }^{\infty }\ \sum\limits_{\lambda =1,2}\left( \hat{k}_{\vec{n}}%
\mathbf{\times }\hat{\varepsilon}_{\vec{n},\lambda }\right) \left[ A_{\vec{n}%
,\lambda }\cos \left( \omega _{\vec{n}}t\right) -B_{_{\vec{n},\lambda }}\sin
\left( \omega _{\vec{n}}t\right) \right] \ \ .  \label{B-in 2}
\end{equation}

The dipole approximation is such that 
\begin{equation}
\cos \left( \vec{k}\cdot \vec{x}-\omega t\right) \approx \cos \left( \omega
t\right)  \label{cosine approx}
\end{equation}%
and 
\begin{equation}
\sin \left( \vec{k}\cdot \vec{x}-\omega t\right) \approx -\sin \left( \omega
t\right) \ \ .  \label{sine approx}
\end{equation}%
This is a good approximation when $\vec{k}\cdot \vec{x}$ $\ll 1$. \ Here, $%
\left\vert \vec{k}\right\vert =\frac{2\pi }{\Lambda }$ where $\Lambda $ is
the wavelength of the light source being considered, is such that $\vec{k}%
\cdot \vec{x}<\left\vert \vec{k}\right\vert \left\vert \vec{x}\right\vert
=2\pi \frac{\left\vert \vec{x}\right\vert }{\Lambda }$ $\ll 1$\ \ . Another
way, and perhaps a better way of applying these approximations that results
in Eqs. (\ref{cosine approx}) and (\ref{sine approx}), is when $\vec{k}\cdot 
\vec{x}\ll \omega t$, which certainly holds for small $\left\vert \vec{x}%
\right\vert $ and large $t$.

Considering first only the electric field in Eq. (\ref{E-in 2}), we can
imagine $\vec{k}_{\vec{n}}$ points in $k-$space as given by Eq. (\ref{kn}).
The value of $A_{\vec{n},\lambda }$ and $B_{_{\vec{n},\lambda }}$ is
\textquotedblleft indexed\textquotedblright\ by the two \textquotedblleft
labels\textquotedblright\ of $\vec{n}$ and $\lambda $. \ The probability
density function in Eq. (\ref{P(A-i)}) describes the Gaussian distribution
of values of $A_{\vec{n},\lambda }$ and $B_{_{\vec{n},\lambda }}$ within the
full ensemble of ZP radiation fields. \ But, just as $A_{\vec{n},\lambda }$
and $B_{_{\vec{n},\lambda }}$ are Gaussian random variables, so are $\left( 
\hat{\varepsilon}_{\vec{n},\lambda }\right) _{i}A_{\vec{n},\lambda }$ and $%
\left( \hat{\varepsilon}_{\vec{n},\lambda }\right) _{i}B_{\vec{n},\lambda }$%
, where $\left( \hat{\varepsilon}_{\vec{n},\lambda }\right) _{i}$ for $%
i=1,2,3$ are the $x,y,z$ components of $\hat{\varepsilon}_{\vec{n},\lambda }$%
.

Suppose we consider all the \textquotedblleft $\vec{n}$\textquotedblright\
points in $k$ 3D space that lie within a shell volume of $k$ space,
described by $4\pi k^{2}dk$. \ Rewriting Eq. (\ref{E-in 2}):%
\begin{equation}
E_{\text{ZP},i}\left( 0,t\right) =\frac{1}{\left( L\right) ^{3/2}}%
\sum\limits _{\substack{ \text{All shells}  \\ \text{of radius }k}}\
\sum\limits_{\vec{n}\ \in \ \text{\textquotedblleft }k\ \text{%
shell\textquotedblright }}\ \sum\limits_{\lambda =1,2}\left( \hat{\varepsilon%
}_{\vec{n},\lambda }\right) _{i}\left[ A_{\vec{n},\lambda }\cos \left(
\omega _{\vec{n}}t\right) -B_{_{\vec{n},\lambda }}\sin \left( \omega _{\vec{n%
}}t\right) \right] \ \ .  \label{E-in 3}
\end{equation}%
In each \textquotedblleft $k$ shell\textquotedblright , $\omega _{\vec{n}}$
is essentially constant for being within the infinitesimal $dk$ of the shell
of volume $4\pi k^{2}dk$. \ Hence we can rewrite Eq. (\ref{E-in 3}) as 
\begin{equation}
E_{\text{ZP},i}\left( 0,t\right) =\sum\limits_{\substack{ \text{All shells} 
\\ \text{of radius }k}}\left\{ 
\begin{array}{c}
\cos \left( \omega _{\vec{n}}t\right) \left[ \frac{1}{\left( L\right) ^{3/2}}%
\sum\limits_{\vec{n}\ \in \ \text{\textquotedblleft }k\ \text{%
shell\textquotedblright }}\ \sum\limits_{\lambda =1,2}\left( \hat{\varepsilon%
}_{\vec{n},\lambda }\right) _{i}A_{\vec{n},\lambda }\right] \\ 
-\sin \left( \omega _{\vec{n}}t\right) \left[ \frac{1}{\left( L\right) ^{3/2}%
}\sum\limits_{\vec{n}\ \in \ \text{\textquotedblleft }k\ \text{%
shell\textquotedblright }}\ \sum\limits_{\lambda =1,2}\left( \hat{\varepsilon%
}_{\vec{n},\lambda }\right) _{i}B_{_{\vec{n},\lambda }}\right]%
\end{array}%
\right\} \ \ .  \label{E-in 4}
\end{equation}

Let 
\begin{equation}
A_{\text{E},i}\left( \omega \right) \equiv \frac{1}{\left( L\right) ^{3/2}}%
\sum\limits_{\vec{n}\ \in \ \text{\textquotedblleft }k\ \text{%
shell\textquotedblright }}\ \sum\limits_{\lambda =1,2}\left( \hat{\varepsilon%
}_{\vec{n},\lambda }\right) _{i}A_{\vec{n},\lambda }  \label{A-Ei}
\end{equation}%
and%
\begin{equation}
B_{\text{E},i}\left( \omega \right) \equiv \frac{1}{\left( L\right) ^{3/2}}%
\sum\limits_{\vec{n}\ \in \ \text{\textquotedblleft }k\ \text{%
shell\textquotedblright }}\ \sum\limits_{\lambda =1,2}\left( \hat{\varepsilon%
}_{\vec{n},\lambda }\right) _{i}B_{_{\vec{n},\lambda }}\ \ ,  \label{B-Ei}
\end{equation}%
recognizing of course $\omega =ck$. \ $A_{\text{E},i}\left( \omega \right) $
and $B_{\text{E},i}\left( \omega \right) $ are Gaussian random variables, as
they are the sum of Gaussian random variables. \ The mean of each is zero,
since $\left\langle A_{\vec{n},\lambda }\right\rangle =\left\langle B_{_{%
\vec{n},\lambda }}\right\rangle =0$\ . \ 

The variance is given by:%
\begin{align}
\left\langle A_{\text{E},i}\left( \omega \right) A_{\text{E},i}\left( \omega
\right) \right\rangle & \equiv \Sigma _{\text{E},i}^{2}\left( \omega \right)
=\left\langle \left[ \frac{1}{\left( L\right) ^{3/2}}\sum\limits_{\vec{n}\
\in \ \text{\textquotedblleft }k\ \text{shell\textquotedblright }}\
\sum\limits_{\lambda =1,2}\left( \hat{\varepsilon}_{\vec{n},\lambda }\right)
_{i}A_{\vec{n},\lambda }\right] \left[ \frac{1}{\left( L\right) ^{3/2}}%
\sum\limits_{\vec{n}^{\prime }\ \in \ \text{\textquotedblleft }k\ \text{%
shell\textquotedblright }}\ \sum\limits_{\lambda ^{\prime }=1,2}\left( \hat{%
\varepsilon}_{\vec{n}^{\prime },\lambda ^{\prime }}\right) _{i}A_{\vec{n}%
^{\prime },\lambda ^{\prime }}\right] \right\rangle  \notag \\
& =\frac{1}{L^{3}}\sum\limits_{\vec{n}\ \in \ \text{\textquotedblleft }k\ 
\text{shell\textquotedblright }}\ \sum\limits_{\lambda =1,2}\ \sum\limits_{%
\vec{n}^{\prime }\ \in \ \text{\textquotedblleft }k\ \text{%
shell\textquotedblright }}\ \sum\limits_{\lambda ^{\prime }=1,2}\left( \hat{%
\varepsilon}_{\vec{n},\lambda }\right) _{i}\left( \hat{\varepsilon}_{\vec{n}%
^{\prime },\lambda ^{\prime }}\right) _{i}\left\langle A_{\vec{n},\lambda
}A_{\vec{n}^{\prime },\lambda ^{\prime }}\right\rangle  \notag \\
& =\frac{1}{L^{3}}\sum\limits_{\vec{n}\ \in \ \text{\textquotedblleft }k\ 
\text{shell\textquotedblright }}\ \sum\limits_{\lambda =1,2}\ \sum\limits_{%
\vec{n}^{\prime }\ \in \ \text{\textquotedblleft }k\ \text{%
shell\textquotedblright }}\ \sum\limits_{\lambda ^{\prime }=1,2}\left( \hat{%
\varepsilon}_{\vec{n},\lambda }\right) _{i}\left( \hat{\varepsilon}_{\vec{n}%
^{\prime },\lambda ^{\prime }}\right) _{i}\delta _{\lambda \lambda ^{\prime
}}\delta _{\vec{n}\vec{n}^{\prime }}2\pi \hbar \omega _{\vec{n}}  \notag \\
& =\frac{1}{L^{3}}\sum\limits_{\vec{n}\ \in \ \text{\textquotedblleft }k\ 
\text{shell\textquotedblright }}\ \sum\limits_{\lambda =1,2}\left( \hat{%
\varepsilon}_{\vec{n},\lambda }\right) _{i}^{2}2\pi \hbar \omega _{\vec{n}}\
\ .  \label{Variance 1E}
\end{align}

However, 
\begin{equation}
\sum\limits_{\lambda =1,2}\left( \hat{\varepsilon}_{\vec{n},\lambda }\right)
_{i}^{2}=1-\left( \frac{k_{i}}{\vec{k}}\right) ^{2}  \label{Eps iden E}
\end{equation}%
Hence,%
\begin{align}
& \left\langle \left[ \frac{1}{\left( L\right) ^{3/2}}\sum\limits_{\vec{n}\
\in \ \text{\textquotedblleft }k\ \text{shell\textquotedblright }%
}\sum\limits_{\lambda =1,2}\left( \hat{\varepsilon}_{\vec{n},\lambda
}\right) _{i}A_{\vec{n},\lambda }\right] \left[ \frac{1}{\left( L\right)
^{3/2}}\sum\limits_{\vec{n}^{\prime }\ \in \ \text{\textquotedblleft }k\ 
\text{shell\textquotedblright }}\sum\limits_{\lambda ^{\prime }=1,2}\left( 
\hat{\varepsilon}_{\vec{n}^{\prime },\lambda ^{\prime }}\right) _{i}A_{\vec{n%
}^{\prime },\lambda ^{\prime }}\right] \right\rangle  \notag \\
& =\frac{1}{L^{3}}\sum\limits_{\vec{n}\ \in \ \text{\textquotedblleft }k\ 
\text{shell\textquotedblright }}\left[ 1-\left( \frac{k_{i}}{\vec{k}}\right)
^{2}\right] 2\pi \hbar \omega _{\vec{n}}\ \ .  \label{Variance 2E}
\end{align}

Since 
\begin{equation*}
\omega _{\vec{n}}=\omega _{\left\vert \vec{n}\right\vert }\text{ }
\end{equation*}%
within this shell of constant $k$, then 
\begin{equation}
\frac{1}{L^{3}}\sum\limits_{\vec{n}\ \in \ \text{\textquotedblleft }k\ \text{%
shell\textquotedblright }}\left[ 1-\left( \frac{k_{i}}{\vec{k}}\right) ^{2}%
\right] 2\pi \hbar \omega _{\vec{n}}=\frac{2\pi \hbar \omega _{\left\vert 
\vec{n}\right\vert }}{L^{3}}\sum\limits_{\vec{n}\ \in \ \text{%
\textquotedblleft }k\ \text{shell\textquotedblright }}\left[ 1-\left( \frac{%
k_{i}}{\vec{k}}\right) ^{2}\right]  \label{w out}
\end{equation}

To evaluate the sum of points within the \textquotedblleft$k$
shell\textquotedblright\ of the quantity $\left[ 1-\left( \frac{k_{i}}{\vec{k%
}}\right) ^{2}\right] $, the sum of any quantity over $N$ points is of
course exactly equal to the average value of that quantity multiplied by $N$%
. \ The question then becomes as to what is the average of $\left[ 1-\left( 
\frac{k_{i}}{\vec{k}}\right) ^{2}\right] $ for points distributed uniformly
around a spherical shell. \ Because of spherical symmetry, the average of $%
\left( k_{x}/k\right) ^{2},$ $\left( k_{y}/k\right) ^{2},$ and $\left(
k_{z}/k\right) ^{2}$ must all be exactly equal. \ Since they must add up to
1, the average value of each one is exactly $1/3$. \ Therefore the average
value of $\left[ 1-\left( \frac{k_{i}}{\vec{k}}\right) ^{2}\right] $ is $%
\frac{2}{3}$. \ Hence:%
\begin{equation}
\sum\limits_{\vec{n}\ \in\ \text{\textquotedblleft}k\ \text{%
shell\textquotedblright}}\left[ 1-\left( \frac{k_{i}}{\vec{k}}\right) ^{2}%
\right] =\frac{2}{3}\times N_{\text{points.}}\ .  \label{avg x N}
\end{equation}

However, for $N_{\text{points.}}=\left( \text{number of points within }%
k\rightarrow k+dk\right) $ , then%
\begin{equation}
N_{\text{points}}=\frac{4\pi k^{2}dk}{\left( \frac{2\pi }{L}\right) ^{3}}=%
\frac{L^{3}}{2\pi ^{2}}\frac{\omega ^{2}d\omega }{c^{3}}\ \ .
\label{Num points}
\end{equation}

Hence, the variance of $A_{\text{E},k,i}$ in Eq. (\ref{A-Ei}) is%
\begin{equation}
\Sigma _{\text{E},i}^{2}\left( \omega \right) =\left\langle A_{\text{E}%
,i}\left( \omega \right) A_{\text{E},i}\left( \omega \right) \right\rangle =%
\frac{2\pi \hbar \omega }{L^{3}}\times \frac{2}{3}\times \frac{L^{3}}{2\pi
^{2}}\frac{\omega ^{2}d\omega }{c^{3}}=\frac{2}{3\pi }\frac{\hbar \omega
^{3}d\omega }{c^{3}}\ \ .  \label{Sigma-2}
\end{equation}%
Thus, this variance is independent of $L$ for large $L$, inversely
proportional to $c^{3}$, and proportional to $\hbar $ and to $\omega ^{3}$.\
\ Also, $\Sigma _{\text{E},i}^{2}\left( \omega \right) $ has the same value
for $i=1,2,3$. \ Moreover, everything above that was done for the variable $%
A_{\text{E},i}\left( \omega \right) $ in Eq. (\ref{A-Ei}) also applies to
the variable $B_{\text{E},i}\left( \omega \right) $ in Eq. (\ref{B-Ei}).

It should be noted that $\left\langle A_{\text{E},i}\left( \omega \right) A_{%
\text{E},j}\left( \omega \right) \right\rangle $ equals zero, for $i\neq j,$
since in the course of deriving it, along the steps above, one obtains $%
\sum\limits_{\lambda =1,2}\left( \hat{\varepsilon}_{\vec{n},\lambda }\right)
_{i}\left( \hat{\varepsilon}_{\vec{n},\lambda }\right) _{j}.$ But, this
equals $-\frac{k_{i}k_{j}}{k^{2}},$ for $i\neq j$. \ When averaged over the
solid angle in $k-$space, this equals zero.

In summary, we can then rewrite Eq. (\ref{E-in 2}) as the following: 
\begin{equation}
E_{\text{ZP},i}\left( 0,t\right) =\sum\limits_{\text{All }j}\left[ A_{\text{E%
},i}\left( \omega _{j}\right) \cos \left( \omega _{j}t\right) -B_{\text{E}%
,i}\left( \omega _{j}\right) \sin \left( \omega _{j}t\right) \right] \ \ ,
\label{E-in 5}
\end{equation}%
where the sum over $j$ is the sum over frequencies, or over the value of $%
\omega _{j}$ in the $j^{\text{th}}$ \textquotedblleft
shell\textquotedblright\ in $k-$space. $A_{\text{E,}i}\left( \omega
_{j}\right) $ and $B_{\text{E,}i}\left( \omega _{j}\right) $ are Gaussian
random variables with mean given by$\left\langle A_{\text{E,}i}\left( \omega
\right) \right\rangle =\left\langle B_{\text{E,}i}\left( \omega \right)
\right\rangle =0$, and variance $\left\langle A_{\text{E,}i}^{2}\left(
\omega \right) \right\rangle =\left\langle B_{\text{E,}i}^{2}\left( \omega
\right) \right\rangle =\Sigma _{\text{E},i}^{2}\left( \omega \right) =\frac{2%
}{3\pi }\frac{\hbar \omega ^{3}d\omega }{c^{3}}$, independent of $i$. \ Each
of the six random numbers $A_{\text{E},i}\left( \omega \right) $ and $B_{%
\text{E},i}\left( \omega \right) $, for $i=1,2,3$ are described by the
probability density distribution%
\begin{equation}
P\left( a\right) =\frac{1}{\sqrt{2\pi \Sigma _{\text{E},i}^{2}\left( \omega
\right) }}\exp \left( -\frac{a^{2}}{2\cdot \Sigma _{\text{E},i}^{2}\left(
\omega \right) }\right) \ \ .  \label{P(v)}
\end{equation}%
\ Thus, $A_{\text{E},i}\left( \omega \right) $ and $B_{\text{E},i}\left(
\omega \right) $ in Eqs. (\ref{A-Ei}) and (\ref{B-Ei}) each represent three
random numbers for $i=1,2,3$, each with a mean of zero and a variance of $%
\Sigma _{\text{E},i}^{2}\left( \omega \right) =\frac{2}{3\pi }\frac{\hbar
\omega ^{3}d\omega }{c^{3}}$, for a total of six independent Gaussian random
numbers. \ 

\subsection{The statistical properties of the ZP magnetic field acting on
the classical electron}

The ZP magnetic field in the dipole approximation appears in Eq. (\ref{Eq
Motion 2}) as $-\frac{e}{c}\frac{d\vec{z}}{dt}\times \vec{B}_{\text{ZP}%
}\left( 0,t\right) $. \ We shall analyze $\vec{B}_{\text{ZP}}\left(
0,t\right) $ in the same way we did for $\vec{E}_{\text{ZP}}\left(
0,t\right) $ in the previous subsection.

From Eq. (\ref{B-in 2})%
\begin{equation}
B_{\text{ZP},i}\left( 0,t\right) =\frac{1}{\left( L\right) ^{3/2}}%
\sum\limits _{\substack{ \text{All shells}  \\ \text{of radius }k}}\
\sum\limits_{\vec{n}\ \in \ \text{\textquotedblleft }k\ \text{%
shell\textquotedblright }}\ \sum\limits_{\lambda =1,2}\left( \hat{k}\times 
\hat{\varepsilon}_{\vec{n},\lambda }\right) _{i}\left[ A_{\vec{n},\lambda
}\cos \left( \omega _{\vec{n}}t\right) -B_{_{\vec{n},\lambda }}\sin \left(
\omega _{\vec{n}}t\right) \right] \ \ .  \label{B-in 3}
\end{equation}%
In each \textquotedblleft $k$ shell\textquotedblright , $\omega _{\vec{n}}$
is essentially constant, for being within the infinitesimal $dk$ of the
shell of volume $4\pi k^{2}dk$. \ Hence we can rewrite Eq. (\ref{B-in 3}) as 
\begin{equation}
B_{\text{ZP},i}\left( 0,t\right) =\sum\limits_{\substack{ \text{All shells} 
\\ \text{of radius }k}}\left\{ 
\begin{array}{c}
\cos \left( \omega _{\vec{n}}t\right) \left[ \frac{1}{\left( L\right) ^{3/2}}%
\sum\limits_{\vec{n}\ \in \ \text{\textquotedblleft }k\ \text{%
shell\textquotedblright }}\ \sum\limits_{\lambda =1,2}\left( \hat{k}\times 
\hat{\varepsilon}_{\vec{n},\lambda }\right) _{i}A_{\vec{n},\lambda }\right]
\\ 
-\sin \left( \omega _{\vec{n}}t\right) \left[ \frac{1}{\left( L\right) ^{3/2}%
}\sum\limits_{\vec{n}\ \in \ \text{\textquotedblleft }k\ \text{%
shell\textquotedblright }}\ \sum\limits_{\lambda =1,2}\left( \hat{k}\times 
\hat{\varepsilon}_{\vec{n},\lambda }\right) _{i}B_{_{\vec{n},\lambda }}%
\right]%
\end{array}%
\right\} \ \ .  \label{B-in 4}
\end{equation}

Let 
\begin{equation}
A_{\text{B},i}\left( \omega \right) \equiv \frac{1}{\left( L\right) ^{3/2}}%
\sum\limits_{\vec{n}\ \in \ \text{\textquotedblleft }k\ \text{%
shell\textquotedblright }}\ \sum\limits_{\lambda =1,2}\left( \hat{k}\times 
\hat{\varepsilon}_{\vec{n},\lambda }\right) _{i}A_{\vec{n},\lambda }
\label{A-Bi}
\end{equation}%
and%
\begin{equation}
B_{\text{B},i}\left( \omega \right) \equiv \frac{1}{\left( L\right) ^{3/2}}%
\sum\limits_{\vec{n}\ \in \ \text{\textquotedblleft }k\ \text{%
shell\textquotedblright }}\ \sum\limits_{\lambda =1,2}\left( \hat{k}\times 
\hat{\varepsilon}_{\vec{n},\lambda }\right) _{i}B_{_{\vec{n},\lambda }}\ \ ,
\label{B-Bi}
\end{equation}%
each being Gaussian random variables. \ The mean of each is again zero,
since $\left\langle A_{\vec{n},\lambda }\right\rangle =\left\langle B_{_{%
\vec{n},\lambda }}\right\rangle =0$. \ The variance of $A_{\text{B},i}\left(
\omega \right) $ is 
\begin{align}
& \left\langle \left[ \frac{1}{\left( L\right) ^{3/2}}\sum\limits_{\vec{n}\
\in \ \text{\textquotedblleft }k\ \text{shell\textquotedblright }%
}\sum\limits_{\lambda =1,2}\left( \hat{k}\times \hat{\varepsilon}_{\vec{n}%
,\lambda }\right) _{i}A_{\vec{n},\lambda }\right] \left[ \frac{1}{\left(
L\right) ^{3/2}}\sum\limits_{\vec{n}^{\prime }\ \in \ \text{%
\textquotedblleft }k\ \text{shell\textquotedblright }}\sum\limits_{\lambda
^{\prime }=1,2}\left( \hat{k}^{\prime }\times \hat{\varepsilon}_{\vec{n}%
,\lambda }\right) _{i}A_{\vec{n}^{\prime },\lambda ^{\prime }}\right]
\right\rangle  \notag \\
& =\frac{1}{L^{3}}\sum\limits_{\vec{n}\ \in \ \text{\textquotedblleft }k\ 
\text{shell\textquotedblright }}\ \sum\limits_{\lambda =1,2}\ \sum\limits_{%
\vec{n}^{\prime }\ \in \ \text{\textquotedblleft }k\ \text{%
shell\textquotedblright }}\ \sum\limits_{\lambda ^{\prime }=1,2}\left( \hat{k%
}\times \hat{\varepsilon}_{\vec{n},\lambda }\right) _{i}\left( \hat{k}%
^{\prime }\times \hat{\varepsilon}_{\vec{n}^{\prime },\lambda ^{\prime
}}\right) _{i}\left\langle A_{\vec{n},\lambda }A_{\vec{n}^{\prime },\lambda
^{\prime }}\right\rangle  \notag \\
& =\frac{1}{L^{3}}\sum\limits_{\vec{n}\ \in \ \text{\textquotedblleft }k\ 
\text{shell\textquotedblright }}\ \sum\limits_{\lambda =1,2}\ \sum\limits_{%
\vec{n}^{\prime }\ \in \ \text{\textquotedblleft }k\ \text{%
shell\textquotedblright }}\ \sum\limits_{\lambda ^{\prime }=1,2}\left( \hat{k%
}\times \hat{\varepsilon}_{\vec{n},\lambda }\right) _{i}\left( \hat{k}%
^{\prime }\times \hat{\varepsilon}_{\vec{n}^{\prime },\lambda ^{\prime
}}\right) _{i}\delta _{\lambda \lambda ^{\prime }}\delta _{\vec{n}\vec{n}%
^{\prime }}2\pi \hbar \omega _{\vec{n}}  \notag \\
& =\frac{1}{L^{3}}\sum\limits_{\vec{n}\ \in \ \text{\textquotedblleft }k\ 
\text{shell\textquotedblright }}\ \sum\limits_{\lambda =1,2}\left( \hat{k}%
\times \hat{\varepsilon}_{\vec{n},\lambda }\right) _{i}^{2}2\pi \hbar \omega
_{\vec{n}}\ \ .  \label{Variance 1B}
\end{align}

However, one can prove that 
\begin{equation}
\sum\limits_{\lambda =1,2}\left( \hat{k}\times \hat{\varepsilon}_{\vec{n}%
,\lambda }\right) _{i}^{2}=1-\left( \frac{k_{i}}{\vec{k}}\right) ^{2}\ \ .
\label{Eps iden B}
\end{equation}%
Hence,%
\begin{align}
& \left\langle \left[ \frac{1}{\left( L\right) ^{3/2}}\sum\limits_{\vec{n}\
\in \ \text{\textquotedblleft }k\ \text{shell\textquotedblright }}\
\sum\limits_{\lambda =1,2}\left( \hat{k}\times \hat{\varepsilon}_{\vec{n}%
,\lambda }\right) _{i}A_{\vec{n},\lambda }\right] \left[ \frac{1}{\left(
L\right) ^{3/2}}\sum\limits_{\vec{n}^{\prime }\ \in \ \text{%
\textquotedblleft }k\ \text{shell\textquotedblright }}\ \sum\limits_{\lambda
^{\prime }=1,2}\left( \hat{k}\times \hat{\varepsilon}_{\vec{n},\lambda
}\right) _{i}A_{\vec{n}^{\prime },\lambda ^{\prime }}\right] \right\rangle 
\notag \\
& =\frac{1}{L^{3}}\sum\limits_{\vec{n}\ \in \ \text{\textquotedblleft }k\ 
\text{shell\textquotedblright }}\left[ 1-\left( \frac{k_{i}}{\vec{k}}\right)
^{2}\right] 2\pi \hbar \omega _{\vec{n}}\ \ .  \label{Variance 2B}
\end{align}

Following now exactly the same set of steps as in the electric field case of
the previous section, from Eqs. (\ref{w out}) to (\ref{Sigma-2}), the
variance in Eq. (\ref{Variance 1B}) can be written as%
\begin{equation}
\Sigma _{\text{B},i}^{2}\left( \omega \right) =\frac{2}{3}\frac{\hbar \omega
^{3}d\omega }{\pi c^{3}}\ \ ,  \label{Sigma-3}
\end{equation}%
which is equal to $\Sigma _{\text{E},i}^{2}\left( \omega \right) $ in Eq. (%
\ref{Sigma-2}). \ The same step above for the variables $A_{\text{B},i}$
apply for the variables $B_{\text{B},i}$.

We can now rewrite Eq. (\ref{B-in 2}) as 
\begin{equation}
B_{\text{ZP},i}\left( 0,t\right) =\sum\limits_{\text{All }j}\left[ A_{\text{B%
},i}\left( \omega _{j}\right) \cos \left( \omega _{j}t\right) -B_{\text{B}%
,i}\left( \omega _{j}\right) \sin \left( \omega _{j}t\right) \right] \ \ ,
\label{B-in 5}
\end{equation}%
where the sum over $j$ is the sum over each frequency $\omega _{j}$ in the $%
j^{\text{th}}$ \textquotedblleft shell\textquotedblright\ in $k-$space. \ As
before, each of the six random numbers $A_{\text{B},i}\left( \omega \right) $
and $B_{\text{B},i}\left( \omega \right) $ depend only on frequency, have
zero mean, and variance $\Sigma ^{2}\left( \omega \right) \equiv \Sigma _{%
\text{E},i}^{2}\left( \omega \right) =\Sigma _{\text{B},i}^{2}\left( \omega
\right) =\frac{2}{3\pi }\frac{\hbar \omega ^{3}d\omega }{c^{3}}$. \ Thus
each of these six \textquotedblleft magnetic field ZP\textquotedblright\
variables are described by the same probability density distribution as
before, namely, Eq. (\ref{P(v)}).

\subsection{Concluding remarks regarding the \textquotedblleft wave-vector
shell\textquotedblright\ analysis}

The statistical distribution of the Lorentz force $-e\vec{E}_{\text{ZP}%
}\left( 0,t\right) -\frac{e}{c}\frac{d\vec{z}}{dt}\times \vec{B}_{\text{ZP}%
}\left( 0,t\right) $ due to the ZP electric and magnetic fields acting on
the classical electron in Eq. (\ref{Eq Motion 2}), imparts its statistical
properties onto the ensemble of electron trajectories via Eqs. (\ref{E-in 5}%
) and (\ref{B-in 5}). \ \ Each of 12 independent random variables, $A_{\text{%
E},i}$, $B_{\text{E},i}$, $A_{\text{B},i}$, $B_{\text{B},i},$ for $i=1,2,3$,
has a mean of zero and variance $\Sigma ^{2}\left( \omega \right) =\frac{2}{%
3\pi }\frac{\hbar \omega ^{3}d\omega }{c^{3}}$. \ Clearly this is enormously
different than needing to provide random numbers for each point in $k$-space
to the $A\,$and $B$ coefficients in Eqs. (\ref{E-in}) and (\ref{B-in}). \
Instead, now for each interval $d\omega $, only 12 random numbers need to be
supplied. \ Of course, the smaller $d\omega $ is made, the more accurate the
stochastic properties are produced. \ To agree with physics in nature,
saving the random number for each $d\omega $ interval is necessary and
results in maintaining the radiation fields throughout time in the
simulation.

Quantifying the exact reduction in memory and execution time provided by
this algorithm is rather complex, where here we are concerned with all the $%
A $ and $B$ variable in Eqs. (\ref{E-in}) and (\ref{B-in}), for $\omega _{%
\text{min}}\approx 10^{14}\frac{\text{rad}}{\text{s}}$ to $\omega _{\text{max%
}}\approx 10^{19}\frac{\text{rad}}{\text{s}}$ (see Sect. 4.1 for these
frequency range estimates). \ However, a rough approximation of the
improvement in data storage requirements is as follows. \ Without the $k-$%
space integration method described from here, memory concerns start with the
volume for a sphere in $k-$space, $\frac{4\pi k^{3}}{3}$, where from Eq. (%
\ref{kn}), 
\begin{equation}
\left\vert \vec{k}_{\vec{n}}\right\vert =\frac{2\pi }{L}\left(
n_{x}^{2}+n_{y}^{2}+n_{z}^{2}\right) ^{1/2}\approx \frac{2\pi }{L}N\mathbf{\
.}  \label{|k mag}
\end{equation}%
Here, $N$ is the approximate number of the $k$ variables along the radius in
the sphere in $k-$space. \ The number of $k-$space points in this sphere,
times four variables at each point, is%
\begin{equation}
\text{\# variables in sphere}=4\times \frac{\left( \frac{4}{3}\pi
k^{3}\right) }{\left( \frac{2\pi }{L}\right) ^{3}}=\frac{16}{3}\pi N^{3}\ \ .
\label{Num var previously}
\end{equation}

In contrast, the number of variables to be stored in this space using the
\textquotedblleft pseudo wave\textquotedblright\ analysis presented here is
approximately given by the following, where $\delta N$ is the number of $k$
points in each $\delta k$ interval of points, where $\delta k=\frac{\delta
\omega }{c}$: 
\begin{equation}
\text{\# variables in }N\text{ \ }k-\text{points}=12\times N\times \frac{%
\delta k}{\delta N}\text{ \ \ .}  \label{Num var pseudo}
\end{equation}%
More precisely, $\delta \omega $ is the interval in Eq. (\ref{Num points}).\
\ Thus the method presented here for the total number of variables to be
stored scales as $N$, where the conventional way above varies as $N^{3}$.

This improvement, along with the log-scaling method to be discussed in the
next section, enables all random variables in the simulations to be stored
totally within the computer's random access memory (RAM), thereby enormously
speeding up retrieval of data for wave integrations carried out at each data
point (twelve times, as will be discussed later \cite%
{Press2007NumericalRecipes}) in the simulation. \ Having the full
integrations done at each point helps to make the ZP radiation Lorentz force
be a smooth function without discontinuities.

\section{Logarithmic scaling and other numerical improvements}

\subsection{Scaling and indexing issues}

Earlier in Sect. 1.0, a range of data was mentioned from radio waves to UV
light, spanning a scale of about $10^{6}$ for \textbf{emitted light }of
frequencies from excited states (in QM terminology) to the ground state in H
atoms. \ In terms of the classical electron orbital frequencies, that the
program actually simulates, this approximate range can be deduced in the
following way. \ For nonrelativistic circular motion, the angular
frequencies of orbits come from balancing the mass times the centripetal
acceleration of the classical H atom to the Coulombic force with the nucleus
(cgs units):%
\begin{equation}
mr\omega ^{2}=\frac{e^{2}}{r^{2}}\ \ ,  \label{Coulombic}
\end{equation}%
giving%
\begin{equation}
\omega =\sqrt{\frac{e^{2}}{mr^{3}}}\ \ .  \label{freq range}
\end{equation}%
This computer program largely covers $r=25\unit{\mathring{A}}$ (although
could readily be extended to much larger $r$ values, but this has been
\textquotedblleft flagged\textquotedblright\ for approximating
\textquotedblleft ionization\textquotedblright ) to $r\approx 0.02\unit{%
\mathring{A}}$ ($v/c$ here is about 4\% the speed of light, so largely in
the nonrelativistic regime). \ This range of orbital radii corresponds to
about $1.3\times 10^{14}\frac{\text{radians}}{\text{s}}$ to $5.7\times
10^{18}\frac{\text{radians}}{\text{s}}$, or representing a span of more than
four decades in orbital frequency.

Using a linear relationship and assuming that the lower frequency range
requires at least 1000 data points to represent each 100$\frac{\text{radian}%
}{\text{s}}$, results in%
\begin{equation}
\text{8 }\frac{\text{bytes}}{\text{word}}\times \frac{\text{12 words}}{\text{%
point}}\times \frac{1000\text{ points}}{100\text{ rad/s}}\times \left( \text{%
Linear range}\right) \approx 6\times 10^{21}\ \text{bytes}\ \ ,
\label{Linear range}
\end{equation}%
for the number of bytes needed to represent the data. \ This occurs because $%
5.7\times 10^{18}\frac{\text{radians}}{\text{s}}-1.3\times 10^{14}\frac{%
\text{radians}}{\text{s}}\approx $ $5.7\times 10^{18}\frac{\text{radians}}{%
\text{s}}$, which represents the approximate physical linear range of data.
\ Clearly $10^{21}$bytes is orders of magnitude beyond what any RAM\ can
hold. \ The situation would be much worse if the 12 words/point were not
made possible by the \textquotedblleft pseudo wave\textquotedblright\ work
in Sec. 3.0, which changed the $k-$space scaling of memory from $O\left(
N^{3}\right) $ to $O\left( N\right) $ scaling.\ Nevertheless, it is still
not enough to deal with the ramifications of Eq. (\ref{Linear range}). \ A
linear array that is dense enough to represent data in the lower frequency
range of $10^{14}\frac{\text{rad}}{\text{s}}$ at $\frac{1000\text{ points}}{%
100\text{ rad/s}}$, when extended to $10^{18}\frac{\text{rad}}{\text{s}}$,
results in the huge number in Eq. (\ref{Linear range}).

By implementing a logarithmic grid, the physical grid spacing $\Delta \omega 
$ is made in proportion to $\omega $. \ This changes the required array size
to the number of decades needed to represent the data, rather than the
absolute linear range of the data. \ When getting to the examples of the use
of our software in Sects. 5.0 and 6.0, this will become even more
meaningful, since log gridding sizes will still be large, but still
acceptable. \ In our most aggressive sizes to date, grids exceeding $10^{8}$
frequency bins will be utilized and discussed. \ Even though this is large,
it is quite acceptable with today's modern computer hardware. \ Two
improvements that make this possible is the reduction in spatial modes in
Sect. 3.0 and the frequency log gridding here in Sect. 4.1.

The logarithmic gridding works is constructed by defining $\omega _{n}$, the
frequency at the $n^{\text{th}}$ point of the division of the spectrum, such
that 
\begin{equation}
\omega _{n+1}=K\omega _{n}\ \ ,  \label{log 3}
\end{equation}%
where $K$ is a constant slightly larger than one. \ Consequently, 
\begin{equation}
\ln \left( \omega _{n+1}\right) =\ln \left( \omega _{n}\right) +\ln \left(
K\right) \ \ .  \label{log 3.5}
\end{equation}%
As the frequencies grow exponentially, their logarithms grow linearly.
Plotting the frequencies on a logarithmic axis results in them being evenly
spaced by a distance of $\ln \left( K\right) $, hence the name
\textquotedblleft logarithmic grid\textquotedblright\ \cite{log-grid-1994}.

From Eq. (\ref{log 3}),%
\begin{equation}
\omega _{n+1}=K^{2}\omega _{n-1}=\ldots =K^{n+1}\omega _{0}\ \ ,
\label{log 4}
\end{equation}%
so that as $n$ grows, so does $\omega _{n}$, but slowly, such that%
\begin{equation}
\frac{\Delta \omega _{n}}{\omega _{n}}=\frac{\omega _{n+1}-\omega _{n}}{%
\omega _{n}}=\frac{\left( K-1\right) \omega _{n}}{\omega _{n}}=\left(
K-1\right) \ \ .  \label{log 5}
\end{equation}%
Thus, $\Delta \omega /\omega $ is a constant. \ As $\omega $ grows via the
many magnitudes of its range, $\Delta \omega $ also grows.

Interestingly enough though, with the advantages in Sec. 3.0 and here in
4.1, memory is no longer the concern. \ Instead, the central processing
unit's (CPU) integration time becomes the main problem. The true penalty of
the linear grid is that numerically computing the Lorentz force in Eq. (\ref%
{Eq Motion 1}) at each time step requires summing over a huge number of
discrete frequency values for the radiation fields, times twelve per
adaptive time step. \ Even with the drastically reduced number of variables
that the pseudo wave method provides, a large bottleneck occurs from a
purely linear grid in terms of the CPU's integration time.

This numerical sum of the radiation field's frequencies at each time step,
times twelve, and over a $\pm 5\%$ window (more on this will be described in
Sect. 4.2) must be carried out. \ At the low end of the spectrum of $1\times
10^{14}\frac{\text{rad}}{\text{s}}$ the divisions $\Delta \omega $ must be
small enough (for example 100 such intervals) to carry out this integration
somewhat accurately. \ However, keeping this same division at the high end
of the spectrum of $6\times 10^{18}\frac{\text{radians}}{\text{s}}$ becomes
approximately 
\begin{equation}
\left( 6\times 10^{18}\frac{\text{radians}}{\text{s}}\right) \cdot \left( 
\frac{100}{1\times 10^{14}\frac{\text{radians}}{\text{s}}}\right) =6\times
10^{6}\ \ ,  \label{log 2}
\end{equation}%
\bigskip integration elements times twelve per time step. \ Thus, time step
integrations become comparatively very slow at the high end of the spectrum.
\ The logarithmic grid described here solves this problem by
\textquotedblleft stretching\textquotedblright\ the bins as $\omega $
increases.

Moreover, suppose $p$ is such that $0<p<1$, then if a \textquotedblleft
window\textquotedblright\ is made of the\ frequencies about $\omega $ of $%
\omega \left( 1-p\right) $ to $\omega \left( 1+p\right) $, as is used in the
algorithm here (Sect. 4.2), then 
\begin{equation}
\ln \left[ \omega \left( 1+p\right) \right] -\ln \left[ \omega \left(
1-p\right) \right] =\ln \left[ \frac{\omega \left( 1+p\right) }{\omega
\left( 1-p\right) }\right] =\ln \left( \frac{1+p}{1-p}\right) \ \ ,
\label{log 6}
\end{equation}%
which is a fixed quantity, only dependent on $p$ and true for all $\omega $
in the spectrum.

Put qualitatively, the frequency spectrum has to span from at least $10^{14}%
\frac{\text{rad}}{\text{s}}$ (the slow outer orbits) all the way to nearly $%
10^{19}\frac{\text{rad}}{\text{s}}$ . If one was to use a pure linear grid
over that entire span, a devastating trade-off would occur: preserving the
necessary $\Delta \omega $ resolution for low frequencies would yield an
astronomical number of bins at the high-frequency regime, paralyzing the CPU
at every time step. The logarithmic scaling method bypasses this trap by
enforcing a constant $\Delta \omega /\omega $ ratio, concentrating
computational resolution only where the physical gradients demand it.

\subsection{Resonance approximation and window algorithm}

Despite the algorithmic improvements of the pseudo waves and logarithmic
grid, the task of accurately computing and simulating the physics of SED is
still impossible without a physically motivated truncation of the frequency
spectrum. \ As can be seen in Eqs. (\ref{E-in}) and (\ref{B-in}) for the ZP\
fields, an infinite sum over the frequencies is present. \ Moreover, the
magnitudes of these terms essentially only increase with frequency, as the
variances of $A_{\vec{n},\lambda }$ and $B_{\vec{n},\lambda }$ are
proportional to $\omega $, as seen in Eqs. (\ref{P(A-i)}) and (\ref{sig-i}).
\ One might then suspect that the SED physics must be \textquotedblleft
wrong.\textquotedblright\ \ Instead, this divergence in SED parallels the
infinities famously tackled by Feynman, Schwinger, and Tomonaga in QED
renormalization. \ Fortunately, for the orbital dynamics of classical
hydrogen in SED, the physical solution is considerably simpler. \ Likely
when examining the fluctuations within an orbit, where the notion of
\textquotedblleft Zitterbewegung\textquotedblright\ starts to apply, then
higher level corrections will be necessary.

The orbital \textquotedblleft solution,\textquotedblright\ at least for the
H atom, was first carried out in 2003 in Ref. \cite{ColeWilsonPRLett2003}
and is repeated here in the present work. The infinite sum over frequencies
is greatly simplified by considering the resonances between the radiation
and the charged particle motion.\ \ (A deeper study is presently underway,
namely, to recognize that the resonance is not just a function of frequency,
but of EM strength as well.) \ Note that it is critical to the present
simulation program that \textquotedblleft windows\textquotedblright\ of
physical resonances do indeed exist between the infinite frequency spectrum
ZP\ radiation and the orbital motion of the classical electron. \ If this
were not true, the present simulation task would be considerably harder.

The method is this: by considering a classical electron in a near circular
orbit in a classical H atom, where the orbit is decaying (spiraling) inward
due to radiation reaction, and by having a monochromatic circularly
polarized (CP) EM wave acting, then a strong resonance is observed when the
orbit's frequency comes close to that of the CP EM wave. Figures 9(a) and
9(b) in Ref. \cite{ColeWilson4th2003} indicate how this was realized. The
orbits of the classical electron in the H atom have a strong response to the
frequencies of the radiation acting on it. At the time of our 2003 and 2004
work, a \textquotedblleft window\textquotedblright\ of $\pm $\ 3\% of the
semimajor axis $a$ of the orbit, which results in a window of $\pm 4.5\%$ in
the net angular frequency, as in Eq. (\ref{freq range}), was found to be
satisfactory. In the simulations reported here, a window of $\pm $\ 5\% has
been used to ensure inclusiveness.

This resonance approach is what Sect. 4.1 was referring to when summing over
the waves within a window [see near Eq. (\ref{log 6})]. \ This resonance
approach means that when evaluating the Lorentz force of $-e\vec{E}_{\text{ZP%
}}\left[ \vec{z}\left( t\right) ,t\right] -\frac{e}{c}\frac{d\vec{z}}{dt}%
\times \vec{B}_{\text{ZP}}\left[ \vec{z}\left( t\right) ,t\right] $ in Eq. (%
\ref{Eq Motion 1}), at time $t$, only the frequencies of $\vec{E}_{\text{ZP}%
} $ and $\vec{B}_{\text{ZP}}$ within $\pm \ 5\%$ of the orbital frequency
need be included.\ This infinity of the ZP\ spectrum is thus dynamically
reduced to a finite numerical summation.

To aid in carrying out this \textquotedblleft sliding
window\textquotedblright\ approach with the orbital frequency $f$, where $%
\omega =2\pi f$, a least squares fit was used to dynamically determine the
semimajor axis length \textquotedblleft $a$\textquotedblright , the
eccentricity $\varepsilon $, and the angle of tilt of the semimajor axis $%
\theta _{0}$, in the following elliptical equation for the radius $r\left(
t\right) $ and angle $\theta \left( t\right) $ of the Coulombic orbit:%
\begin{equation}
r\left( t\right) =\frac{a\left( 1-\varepsilon ^{2}\right) }{1-\varepsilon
\cos \left[ \theta \left( t\right) -\theta _{0}\right] }\ \ .
\label{least sq}
\end{equation}

If the ZP radiation did not exist, nor radiation reaction, then Eq. (\ref%
{least sq}) would be the correct elliptical orbit of the electron. \ With ZP
radiation and radiation reaction present, however, then Eq. (\ref{least sq})
is only approximately valid, but reasonably close to the actual orbit for a
number of orbits, due to the strength of the H nucleus acting. \
Consequently, $a$, $\varepsilon $, and $\theta _{0}$ are all slowly varying
functions of time $t.$ These parameters were obtained by rewriting Eq. (\ref%
{least sq}) as 
\begin{equation}
\frac{1}{r}=\frac{1}{a\left( 1-\varepsilon ^{2}\right) }-\left[ \frac{%
\varepsilon \cos \theta _{0}}{a\left( 1-\varepsilon ^{2}\right) }\right]
\cos \theta -\left[ \frac{\varepsilon \sin \theta _{0}}{a\left(
1-\varepsilon ^{2}\right) }\right] \sin \theta \ \ .  \label{least sq 2}
\end{equation}%
By making a table of $1/r$, $\cos \theta $, and $\sin \theta $, for $N$
points of an orbit or orbits, the parameters $\frac{1}{a\left( 1-\varepsilon
^{2}\right) }$, $-\left[ \frac{\varepsilon \cos \theta _{0}}{a\left(
1-\varepsilon ^{2}\right) }\right] ,$ and $-\left[ \frac{\varepsilon \sin
\theta _{0}}{a\left( 1-\varepsilon ^{2}\right) }\right] $ were obtained
using the least squares method since these parameters appear linearly in the
above relationship. \ From these three extracted parameters, $a$, $%
\varepsilon $, and $\theta _{0}$ were obtained.

Upon finding $a$, then via Kepler's third law, one obtains the period $T_{0}$%
, and then the orbital frequency is deduced by $f=\frac{1}{T_{0}}$. \ This
least squares fit method was developed and published in Ref. \cite%
{ColeWilson2nd2002}. \ This method was used here to obtain a close track of
the net (central) angular frequency $\omega _{c}=2\pi f$, and therefore on
the sliding window frequencies to be used in the Lorentz force summation of
the ZP\ fields. \ It is this summation over frequencies that is carried out
at each integration point (12 times) in the orbits.

This last point brings us back to the use and advantages of the logarithmic
gridding system in Sect. 4.1. \ The exact number of frequencies to be summed
over in the $\pm 5\%$ window depends on the total grid size $N_{\text{total}%
} $, which means the total number of discrete frequencies in the overall
simulation. Because of the logarithmic scale, the $\pm 5\%$ window always
contains a fixed fraction of the $N_{\text{total}}$ array, roughly 0.94\%. \
More specifically, the total spectrum spans from $\omega _{\text{min}%
}=1.3\times 10^{14}$ to $\omega _{\text{max}}=5.7\times 10^{18}\frac{\text{%
rad}}{\text{s}}$. \ $\ln \left( \frac{5.7\times 10^{18}}{1.3\times 10^{14}}%
\right) \approx 10.688.$ \ The sliding window spans from $0.95\omega _{c}$
to $1.05\omega _{c}$. $\ln \left( \frac{1.05}{0.95}\right) \approx 0.1001$.
In terms of the logarithmic gridding, the fraction of the number of
frequencies inside the sliding window to the fraction of the net wave
frequency span is%
\begin{equation*}
\frac{\ln \left( \frac{1.05}{0.95}\right) }{\ln \left( \frac{5.7\times
10^{18}}{1.3\times 10^{14}}\right) }\approx 0.0094\ \ \text{. \ }
\end{equation*}%
\bigskip Therefore, the $\pm 5\%$ window always captures approximately
0.94\% of the total grid, or the total number of discrete frequencies used
in the frequency range from $\omega _{\text{min}}$ to $\omega _{\text{max}}$.

Putting some numbers on this, if there are 1M discrete frequencies (bins) in
the simulation, the CPU performs the sum of 9,360 Lorentz force $\vec{F}_{%
\text{Lor}}$ values per time step, times three, since $\vec{F}_{\text{Lor}}$
is a vector in 3-space, and times 12 since the RK step-doubling method is
used. \ If 10M net discrete frequencies are used, then 93,600 discrete
frequencies of the Lorentz force is summed over, and with an aggressive $100$%
M bins, the most tried to date, this balloons to 936,000 per time step,
times the previous factors. \ Despite being large, the $\pm 5\%$ window
restricts the CPU to computing less than 1\% ($\approx $0.94\%) of the total
domain at any given integration step, which helps make the simulation
computationally tractable.

\subsection{Smoothing function and related discontinuity avoidances}

The sliding window described in the previous section is based on the
approximate resonance width of the nearly elliptical orbit interacting with
the ZP radiation. However, applying a strict rectangular truncation---where
frequencies within the interval $\omega _{c}\left( 1-p\right) $ to $\omega
_{c}\left( 1+p\right) $ contribute strongly to the Lorentz force, while
frequencies immediately outside contribute nothing---introduces an
artificial discontinuity. \ Physically, the influence of the radiation
spectrum on the electron's resonance cannot fall off instantaneously. That
makes little sense. \ Moreover, computationally, a hard rectangular boundary
guarantees that whenever a discrete wave enters or exits the sliding window
during integration, the differential equation solver will encounter a
severe, non-physical step-change in the calculated Lorentz force.

If the discontinuity is allowed for either reason, the effect will be the
same. \ If a wave amplitude instantly drops from 1 to 0 at exactly the
frequency \textquotedblleft window\textquotedblright\ edge, the derivative
of the Lorentz force will mathematically spike to infinity. The adaptive RK4
step-doubling algorithm will interpret this as a massive local truncation
error; it will \textquotedblleft halve\textquotedblright\ the time step
until it hits the minimum time step set in the integration software, which
either results in the software aborting or getting trapped in an inescapable
bottleneck.

To resolve this, a smoothing function was introduced to render the frequency
truncation mathematically continuous. The cosine-tapered Tukey window \cite%
{BlackmanTukey1958}, \cite{Harris1978} provides exactly this capability. By
applying this function, the amplitudes of the contributing EM waves are
smoothly \textquotedblleft tapered\textquotedblright\ to zero at the edges
of the frequency interval. This not only captures the physical reality of a
continuous resonance drop-off, but also provides a vastly superior numerical
capability for the integration algorithm. \ The \textquotedblleft
window\textquotedblright\ of the frequency modifies the wave's amplitude
contribution by the following factor, as a function of $\omega $:

\begin{equation}
W(\omega )=%
\begin{cases}
\frac{1}{2}\left[ 1+\cos \left( \pi \left( \frac{\omega -\omega _{1}}{\Delta
_{\text{tapper}}}-1\right) \right) \right] & \omega _{\text{min}}\leq \omega
<\omega _{1} \\ 
1 & \omega _{1}\leq \omega \leq \omega _{2} \\ 
\frac{1}{2}\left[ 1+\cos \left( \pi \left( \frac{\omega -\omega _{2}}{\Delta
_{\text{tapper}}}\right) \right) \right] & \omega _{2}<\omega \leq \omega _{%
\text{max}} \\ 
0 & \text{otherwise}%
\end{cases}%
\ \ ,  \label{W(w)}
\end{equation}%
where $\Delta _{\text{taper}}$ is the frequency width of the
fade-in/fade-out regions, $\omega _{1}=\omega _{\text{min}}+\Delta _{\text{%
taper}}$ is where the flat top begins,$\ $and $\omega _{2}=\omega _{\text{max%
}}-\Delta _{\text{taper}}$ is where the flat top ends and the fade-out
begins. \ Thus, the center of the above function provides a flat 1.0
amplitude multiplier for $\omega _{1}\leq \omega \leq \omega _{2}$, while
the edges use a cosine curve to provide a smooth continuous transition to
zero. \ The taper width can be numerically experimented with to achieve the
best numerical stability. The control of this function in the present
software will be clarified further in Sect. 5.0.

Another source of discontinuities and likely severe computational barrier is
the following. \ As briefly mentioned in Sect. 4.2, a tempting computational
shortcut is to evaluate the EM wave summation only once or a few times per
orbit. Once the initial grid is established and all related physical
parameters are stored (\textit{i.e.,} $A^{\prime }s$ and $B^{\prime }s$
values in Sect. 3.0, 12 for each bin or $\Delta \omega $ frequency
interval), the vast majority of the simulation's CPU time is spent carrying
out these summations. Hence, only updating the sum every $N^{\text{th}}$
integration step might seem a needed and reasonable computational
approximation.

However, over such a \textquotedblleft macroscopic\textquotedblright\
duration, the instantaneous phases and fields change too rapidly. \ Updating
the sum so infrequently can introduce large discrete jumps in the computed
Lorentz force, violently kicking the electron and generating non-physical
spikes in the orbital trajectory. \ Such artificial numerical shocks will
likely lead to false trajectory predictions, including premature orbital
decays or spurious ionizations. \ Quite possibly the ionization events
reported in Refs. \cite{Nieuwenhuizen2015-Nonrel}, \cite%
{Nieuwenhuizen2015-Relativistic} were at least partially caused in this way.
\ After all, the eccentricity of the orbit reported in Ref. \cite%
{Nieuwenhuizen2015-Nonrel} was highly spiked, with values peaking as high as
0.8. \ For an ellipse, $0\leq \varepsilon <1$, with $\varepsilon \approx 0$
being a circle. \ In contrast, most of our simulations to date have been
yielding orbits close to being circular and not highly elliptic at all. \ 
\textit{I.e.}, see panel \#3 in Fig. 1. \ In hindsight, this result of
nearly circular orbits seems physically plausible since ZP radiation is
homogeneous and isotropic; consequently it \textquotedblleft
bombards\textquotedblright\ the electron from all sides, over and over again.

\subsection{Need for Runge-Kutta adaptive time step algorithm}

The RK4 algorithm with \textquotedblleft step-doubling\textquotedblright\
(Richardson Extrapolation) serves as the primary numerical workhorse for the
simulations presented here. While methods such as Bulirsch-Stoer or
predictor-corrector algorithms can be highly efficient for certain systems,
they struggle when evaluating the highly stochastic, rapidly fluctuating
electromagnetic fields inherent to this model \cite%
{Press2007NumericalRecipes}.\ Bulirsch-Stoer, for example, relies on
rational function extrapolation assuming a smooth underlying field, making
it unsuitable for the \textquotedblleft ragged\textquotedblright\
environment of the ZP radiation.

The step-doubling approach relies on the standard four-stage RK4 evaluation,
but executes this three separate times (12 total evaluations) to compare
different step sizes. Step-doubling is exceptionally robust for highly
fluctuating trajectories because the two half-steps provide a highly
accurate final state, but it is computationally heavier than the embedded
Runge-Kutta-Fehlberg (RKF45) method.

As discussed in the previous subsection, discontinuous step-changes in the
evaluated forces can introduce severe instabilities into the integrator's
predictions---a vulnerability shared by all differential equation solvers.
The implementation of the continuous smoothing functions detailed in Sect.
4.3 directly mitigates this, providing the RK4 solver with a continuously
differentiable framework.

The adaptive step algorithm used here plays an important role when the size
of the orbit shrinks or grows, by saving steps when not needed and
vice-versa. \ It is essential when considering an elliptical orbit,
particularly when the orbit has a high eccentricity $\varepsilon $. \ The
speed near the nucleus (periapsis) is then far larger than the speed of the
electron at the extreme distance from the nucleus (apoapsis). \ It is not
uncommon to require, for example, 20,000 integration points near the nucleus
compared to 1,000 or fewer at the apoapsis.

To maintain accuracy across vastly different orbital regimes such as these,
the adaptive step algorithm employs an error estimate. \ The algorithm
adjusts the step size to meet this criterion. \ The result is that the
adaptive time-step algorithm dynamically scales the integration increment to
meet strict error tolerances without unnecessary computational overhead. \
With the computational demands of the core physics involving a rapidly
varying Lorentz ZP force, both in time and space, this RK4 algorithm with
step doubling is essential for carrying out the computational aspects
detailed here.

\section{Development analysis and validation}

\subsection{Overview}

The development of a robust simulation for a classical electron in a
hydrogenic potential, bathed in zero-point (ZP) electromagnetic radiation,
presents a unique validation challenge: there are no exact intermediate
analytical expressions for this specific system to benchmark against. The
ultimate test of the code is solely whether the final SED predictions
reproduce QM. Consequently, establishing rigorous intermediate validation
checks during code development required leveraging related, solvable
analytical systems.

This section goes over the primary checks used to confirm the physics and
algorithms of the simulation engine. Subsection 5.1 compares the
simulation's generated ZP fields against exact analytical results for the
two-point spatial and temporal correlation functions. Subsection 5.2
benchmarks the simulation using the SHO by comparing the numerically
generated variance of a charged particle's position against a known
analytical SED expression. Finally, Subsections 5.3 and 5.4 provide a
computational analysis justifying the inclusion or omission of specific
terms within the equation of motion, Eq. (\ref{Eq Motion 1}). Together,
these analyses confirmed the numerical stability and physical accuracy of
the underlying architecture before applying it to the full hydrogen atom.

\subsection{Comparison with two-point ZP field correlation functions}

During the development of the 3D simulations, it was critical to verify that
the discrete sum over the ZP modes using Gaussian-generated random
amplitudes [see Eqs. (\ref{E-in}), (\ref{B-in}), (\ref{P(A-i)})] correctly
replicates the continuous physical vacuum. To test this, we evaluated
several examples of the ensemble average of the two-point spatial-temporal
ZP field correlation functions $\langle E_{\text{ZP,}i}(t_{1},\vec{x}_{1})E_{%
\text{ZP,}j}(t_{2},\vec{x}_{2})\rangle $ and $\langle E_{\text{ZP,}i}(t_{1},%
\vec{x}_{1})B_{\text{ZP,}j}(t_{2},\vec{x}_{2})\rangle $. \ One of these
examples will be cited here, namely, of $\langle E_{\text{ZP,}x}(0,0)E_{%
\text{ZP,}x}(t,R\hat{y})\rangle $, to validate the stochastic field
generation engine. 

The code executes the following steps to ensure accuracy:

\begin{itemize}
\item Analytical Benchmarking: The program establishes two theoretical
baselines. First, it calculates the exact closed-form analytical solution
for an infinite spectrum from $\omega =0$ to $\infty $. Second, it computes
a finite-bandwidth continuous integral (from $\omega _{\text{min}}$ to $%
\omega _{\text{max}}$) to account for the exact numerical truncation bounds
used in the simulation.

\item Logarithmic Grid Geometry (Precomputation): The engine initializes a
logarithmic frequency grid composed of 10,000 discrete modes. For each
frequency bin, it precalculates an isotropic random wave-vector direction ($%
\hat{k}$) and an orthogonal polarization plane, projecting this geometry to
extract the polarization vector directions. It also establishes the
zero-point standard deviation ($\sigma \propto \sqrt{\hbar \omega }$) for
each individual mode.

\item Monte Carlo Ensemble Generation: The simulation iterates over 10,000
distinct statistical universes (ensembles). For every frequency mode within
each universe, it draws random Gaussian Fourier coefficients ($A\ $and $B$)
governed by the precomputed $\sigma $ variance.

\item Field Evaluation and Density Scaling: In each ensemble, the code
measures the simulated ZP fields at two locations: the spatial origin at $t=0
$ and $\vec{x}=0$ and a target atomic distance $R$ at time $t$. Critically,
the amplitude of each mode is multiplied by a 3D density-of-states weighting
factor. This specific mathematical scaling dynamically maps the 1D
logarithmic grid back into the physical 3D continuum.

\item Convergence and Comparison: The program multiplies the two ZP field
values, accumulates them across the 10,000 ensembles, and calculates the
final numerical average. This value is then output alongside the theoretical
targets to demonstrate whether the randomized algorithmic approach
accurately reproduces the exact analytical physics.
\end{itemize}

\begin{center}
$%
\begin{tabular}{|c|c|c|c|}
\hline
Separation $R$ (A) & Analytical $\omega _{\text{max}}\rightarrow \infty $ & 
Analytical (Finite $\omega _{\text{max}}$) & C++ Numerical Average (10k
ensembles) \\ \hline
0.1 & $7.4544\times 10^{18}$ & $2.0764\times 10^{8}$ & $1.9351\times 10^{8}$
\\ \hline
1.0 & $7.4544\times 10^{14}$ & $2.0732\times 10^{8}$ & $2.0732\times 10^{8}$
\\ \hline
10.0 & $7.4544\times 10^{10}$ & $1.7640\times 10^{8}$ & $1.7297\times 10^{8}$
\\ \hline
\end{tabular}%
$

\textbf{Table 1:} Three ways of evaluating $\langle E_{\text{ZP,}x}(0,0)E_{%
\text{ZP,}x}(t,R\hat{y})\rangle $. \ Columns 2 and 3 represent the
conventional method with a fixed integrand format, evaluated to an infinite
limit and a finite cutoff $k_{\text{max}}=\omega _{\text{max}}/c$,
respectively.. \ Column 4 is evaluated as in this simulation program, where
the integrand at each frequency bin is generated by a Gaussian random number
according to Eq. (\ref{P(A-i)}). \ Columns 3 and 4 are compared, where the
integrand is taken over a finite maximum $k_{\text{max}}=\omega _{\text{max}%
}/c$ value. \ The second, third, and fourth columns all have units of $%
\left( \frac{\text{statvolt}}{\text{cm}}\right) ^{2}$ \ .\bigskip 
\end{center}

From Ref. \cite{Cole1987}, Appendix C, the analytical form is given by: 
\begin{eqnarray}
\langle E_{\text{ZP,}x}(0,0)E_{\text{ZP,}x}(t,R\hat{y})\rangle  &=&\frac{1}{2%
}\dint d^{3}k\frac{\hbar \omega }{2\pi ^{2}}\left( 1-\frac{k_{x}^{2}}{k^{2}}%
\right) \cos \left( k_{y}R-kct\right)   \notag \\
&=&\frac{\hbar c}{\pi }\dint_{0}^{k_{\text{max}}}k^{3}dk\left[ \frac{\sin
\left( kR\right) }{kR}+\frac{\cos \left( kR\right) }{\left( kR\right) ^{2}}-%
\frac{\sin \left( kR\right) }{\left( kR\right) ^{3}}\right] \cos \left(
kct\right)   \notag \\
&&\underset{k_{\text{max}}\rightarrow \infty }{\rightarrow }+\frac{4\hbar c}{%
\pi }\left\{ \frac{R^{2}+\left( ct\right) ^{2}}{\left[ \left( ct\right)
^{2}-R^{2}\right] ^{3}}\right\} \ \ .  \label{ExEx}
\end{eqnarray}

The second column in Table 1 is given by the third line of Eq. (\ref{ExEx});
the third column is given by the second line of Eq. (\ref{ExEx}), as
evaluated in Mathematica. The fourth column is given by our simulator that
sums up the random amplitudes at each frequency, weighted by the density of
states, to construct the total electric field at the spatial origin at time $%
t=0$. \ The simulator does the same thing at position $R\hat{y}$ and time $t$%
.\ Adding the products up in $E_{\text{ZP,}x}(0,0)E_{\text{ZP,}x}(t,R\hat{y})
$, over 10,000 bins, yields column 4. \ As shown, the finite-bandwidth
analytical results (Column 3) and the simulation output (Column 4) are in
excellent agreement. \ 

This validation procedure was successfully repeated for four different
correlation functions, such as: 
\begin{equation*}
\left\langle E_{\text{ZP,}x}(0,0)B_{\text{ZP,}z}(t,R\hat{y})\right\rangle 
\underset{k\rightarrow \infty }{\rightarrow }-\frac{4\hbar c}{\pi }\left\{ 
\frac{2Rct}{\left[ \left( ct\right) ^{2}-R^{2}\right] ^{3}}\right\} \ \ .
\end{equation*}%
Similar agreement as in Table 1 were found for these different correlation
functions.

\subsection{Use of simulator to \textquotedblleft
randomly\textquotedblright\ evaluate $\left\langle r^{2}\right\rangle $ for
the SHO}

During the course of development, a further check was made on how well the
summation over random generated modes, weighted by the ZP spectrum frequency
functional value, agreed with an analytic expression. \ Instead of just a
check on the fields $\vec{E}_{\text{ZP}}$ and $\vec{B}_{\text{ZP}}$, as in
the previous subsection, here the effect of these fields was evaluated for
the behavior of the classical electron in a SHO binding potential. The
reason for choosing this case is that known analytic solutions existed for
the SHO that could be used in making such a comparison. \ In particular, the
ensemble average of the square of the radius $\left\langle
r^{2}\right\rangle $ for a classical electron charge in a SHO binding
potential can be evaluated analytically as well as by simulation. \ 

The equation of motion examined was%
\begin{equation}
m\frac{d^{2}}{dt^{2}}\vec{z}\left( t\right) =-m\omega _{0}^{2}\vec{z}\left(
t\right) -m\tau \omega _{0}^{2}\frac{d}{dt}\vec{z}\left( t\right) -e\vec{E}_{%
\text{zp}}\left( 0,t\right) \ \ .  \label{Eq Motion SHO}
\end{equation}%
This equation is a linear ordinary differential equation (ODE) that is also
a stochastic one, because of the presence of the \textquotedblleft driving
term\textquotedblright\ $-e\vec{E}_{\text{zp}}\left( 0,t\right) $. \ It
should be contrasted with the expression in Eqs. (\ref{Eq Motion 2}) and (%
\ref{Frr}). \ In Eq. (\ref{Eq Motion SHO}) above, $-m\omega _{0}^{2}\vec{z}%
\left( t\right) $ is the binding force for the electron in a SHO binding
potential $\frac{m\omega _{0}^{2}\vec{z}^{2}}{2}$, with $\omega _{0}$ being
the resonant frequency of the oscillator. \ The radiation reaction $\vec{F}_{%
\text{rr}}$ expression of $-m\tau \omega _{0}^{2}\frac{d}{dt}\vec{x}\left(
t\right) $ came from approximating Eq. (\ref{Frr}) via 
\begin{equation}
m\tau \frac{d^{3}\vec{z}}{dt^{3}}\approx m\tau \frac{d}{dt}\left( -m\omega
_{0}^{2}\vec{z}\left( t\right) \right) \ \ ,  \label{Frr SHO 1}
\end{equation}%
following the same reasoning as in Eqs. (\ref{Frr Coul 2}) and (\ref{Frr
Coul 3}).

The other major difference between Eq. (\ref{Eq Motion 2}) and Eq. (\ref{Eq
Motion SHO}) is that the term 
\begin{equation*}
-\frac{e}{c}\frac{d\vec{z}}{dt}\times \vec{B}_{\text{ZP}}\left( 0,t\right)
\end{equation*}%
was dropped due to the assumption that 
\begin{equation}
\frac{1}{c}\left\vert \frac{d\vec{z}}{dt}\right\vert \ll 1\ \ .  \label{v/c}
\end{equation}

As discussed in many SED references such as Ref. \cite%
{ColePhysicsTwo-Methods-2023} [see Eq. (36), times 3 because of three
degrees of freedom]:%
\begin{equation}
\left\langle r^{2}\left( t\right) \right\rangle \approx 3\frac{(e/m)^{2}}{%
\left( 2\pi \right) ^{3}}\frac{8\pi }{3}\frac{1}{c^{3}}\dint\limits_{0}^{%
\infty }d\omega \frac{\omega ^{2}\sigma ^{2}\left( \omega ,T\right) }{\left(
-\omega ^{2}+\omega _{0}^{2}\right) ^{2}+\left( \tau \omega ^{3}\right) ^{2}}%
\ \ .  \label{Semi-Anal}
\end{equation}%
where $T=0$ for the ZP case. \ In the limit of $\tau \rightarrow 0$, and
with $\sigma ^{2}\left( \omega ,T=0\right) =2\pi \hbar \omega $, this becomes%
\begin{equation}
\left\langle r^{2}\left( t\right) \right\rangle \rightarrow \frac{3\hbar }{%
2m\omega _{0}}\ \ ,  \label{r2 2}
\end{equation}%
a familiar expression for the expectation value of the variance of the 3D
SHO position in QM. \ It should be noted that Eq. (\ref{Semi-Anal}) was
obtained by writing the steady state solution of $\vec{z}\left( t\right) $
in Eq. (\ref{Eq Motion SHO}) as a function of $\vec{E}_{\text{ZP}}\left(
0,t\right) $ in Eq. (\ref{E-in}), which in turn depends on the random
variables of $A^{\prime }s$ and $B^{\prime }s$, and then taking the ensemble
average $\left\langle \vec{z}\cdot \vec{z}\right\rangle $, using the
properties of the $A^{\prime }s$ and $B^{\prime }s$ in Eqs. (\ref{<A>})-(\ref%
{<AA'>}).

What was helpful about all of this in the development of the present
simulation program is that Eq. (\ref{Semi-Anal}) can be evaluated in two
ways, one via analytically carrying out the integral via numerical means
(such as via Mathematica), or by using the simulation program to numerically
solve Eq. (\ref{Eq Motion SHO}) over a length of time, assuming ergodicity,
and finding the time average of 
\begin{equation}
\left\langle r^{2}\left( t\right) \right\rangle \approx \frac{1}{T_{\text{m}}%
}\int_{0}^{T_{\text{m}}}dt\left\vert \vec{z}\left( t\right) \right\vert
^{2}\ \ ,  \label{r^2 Tm}
\end{equation}%
for large values of $T_{\text{m}}$ , and where the numerical values of $%
m=9.1094\times 10^{-28}$g, $e=4.8032\times 10^{-10}$esu, and $\omega
_{0}=4.5\times 10^{16}$s$^{-1}$(frequency close to that of the Bohr orbit)
were used. \ Instead of $\tau =\frac{2e^{2}}{3mc^{3}}=6.2660\times 10^{-24}\ 
$s, the values of $f\times 6.2660\times 10^{-24}$\ s were inserted, where
different values of $f$ were used for testing and comparing Eqs. (\ref%
{Semi-Anal}) and (\ref{r^2 Tm}), as seen in the table below. \ For Eq. (\ref%
{r^2 Tm}), $T_{\text{m}}=1\ $ps was used. \ As can be seen below, columns 2
and 3, representing the two methods, agreed quite well over this large range
of values for $f$.

\begin{center}
\ $%
\begin{tabular}{|c|c|c|}
\hline
$f$ & Mathematica result $\left( \unit{cm}^{2}\right) $ of Eq. (\ref%
{Semi-Anal}) & C++ Simulator result $\left( \unit{cm}^{2}\right) $ of Eq. (%
\ref{r^2 Tm}) \\ \hline
$10^{5}$ & $4.064\times 10^{-22}$ & $3.814\times 10^{-22}$ \\ \hline
$10^{4}$ & $3.895\times 10^{-21}$ & $3.824\times 10^{-21}$ \\ \hline
$10^{3}$ & $3.863\times 10^{-20}$ & $3.968\times 10^{-20}$ \\ \hline
\end{tabular}%
$

\textbf{Table 2: }Comparison of calculating $\left\langle r^{2}\left(
t\right) \right\rangle $ via the analytic Eq. (\ref{Semi-Anal}) and the
simulated method of Eq. (\ref{r^2 Tm}).
\end{center}

\subsection{Check on ZP Lorentz force inclusion in the radiation reaction}

The development of the simulation program described here included an
examination of physical effects that should be included in Eq. (\ref{Eq
Motion 1}). \ Two main areas in particular were explored: (1) examining
including the contribution of $-e\vec{E}_{\text{ZP}}\left[ \vec{z}\left(
t\right) ,t\right] -\frac{e}{c}\frac{d\vec{z}}{dt}\times \vec{B}_{\text{ZP}}%
\left[ \vec{z}\left( t\right) ,t\right] $ to the radiation reaction, and (2)
the dipole approximation, which Sect. 3.0 (pseudo wave analysis) was
partially based upon.

Regarding (1), the standard procedure in SED is to have the radiation
reaction depend only on the Coulombic force, as in Eqs. (\ref{Frr Coul 2})-(%
\ref{Frr Coul 4}). \ However, it is natural to ask about using%
\begin{equation}
m\tau \frac{d^{3}\vec{z}}{dt^{3}}\approx \tau \frac{d}{dt}\left( -\frac{e^{2}%
}{\left\vert \vec{z}\right\vert ^{3}}\hat{z}-e\vec{E}_{\text{ZP}}\left[ \vec{%
z}\left( t\right) ,t\right] -\frac{e}{c}\frac{d\vec{z}}{dt}\times \vec{B}_{%
\text{ZP}}\left[ \vec{z}\left( t\right) ,t\right] \right) \ \ .
\label{Frr 01}
\end{equation}%
To examine this effect of the above extra terms in the radiation reaction
expression, the dipole approximation was made in the fields $\vec{E}_{\text{%
ZP}}$ and $\vec{B}_{\text{ZP}}$ and the expressions of Eqs. (\ref{E-in 2})
and (\ref{B-in 2}) were used. \ However, the sum over all the modes in (\ref%
{E-in 2}) and (\ref{B-in 2}) was changed to include only the waves that were
used in the equation of motion, for each value of $\omega $, due to the
\textquotedblleft window resonance\textquotedblright\ discussed in Sect. 4.2.

Starting the simulation with the classical electron in a circular orbit of
radius $0.53\unit{\mathring{A}}$, and using Eq. (\ref{Frr 01}), with the
restrictions just mentioned, the magnitude 
\begin{equation*}
\left\vert \tau \frac{d}{dt}\left( \frac{e^{2}\hat{z}}{\left\vert \vec{z}%
\right\vert ^{3}}\right) \right\vert \ \ ,
\end{equation*}%
which represents the Coulombic contribution to the radiation reaction, was
plotted, as was the magnitude%
\begin{equation*}
\left\vert \tau \frac{d}{dt}\left( -e\vec{E}_{\text{ZP}}\left[ \vec{z}\left(
t\right) ,t\right] -\frac{e}{c}\frac{d\vec{z}}{dt}\times \vec{B}_{\text{ZP}}%
\left[ \vec{z}\left( t\right) ,t\right] \right) \right\vert \ \ ,
\end{equation*}%
with the restrictions just mentioned, also included. \ Figure 1 shows the
result, with the y-axis using a log scale to show the large difference in
the magnitudes. \ As can be seen, the ZP contribution to $\vec{F}_{\text{rr}}
$ from the ZP Lorentz EM force is orders of magnitude smaller than the
contribution of the Coulombic to $\vec{F}_{\text{rr}}$. Hence, the
approximation of 
\begin{equation}
\vec{F}_{\text{rr}}\approx \tau \frac{d}{dt}\left( \frac{e^{2}\hat{z}}{%
\left\vert \vec{z}\right\vert ^{3}}\right) =\tau e^{2}\left[ -\frac{\frac{d%
\vec{z}}{dt}}{\left\vert \vec{z}\right\vert ^{3}}+\frac{3\vec{z}\mathbf{(}%
\vec{z}\mathbf{\cdot }\frac{d\vec{z}}{dt})}{\left\vert \vec{z}\right\vert
^{5}}\right] \ \ ,  \label{Frr Coulomb}
\end{equation}%
is an excellent one.

\begin{center}
\includegraphics[width=0.9\textwidth]{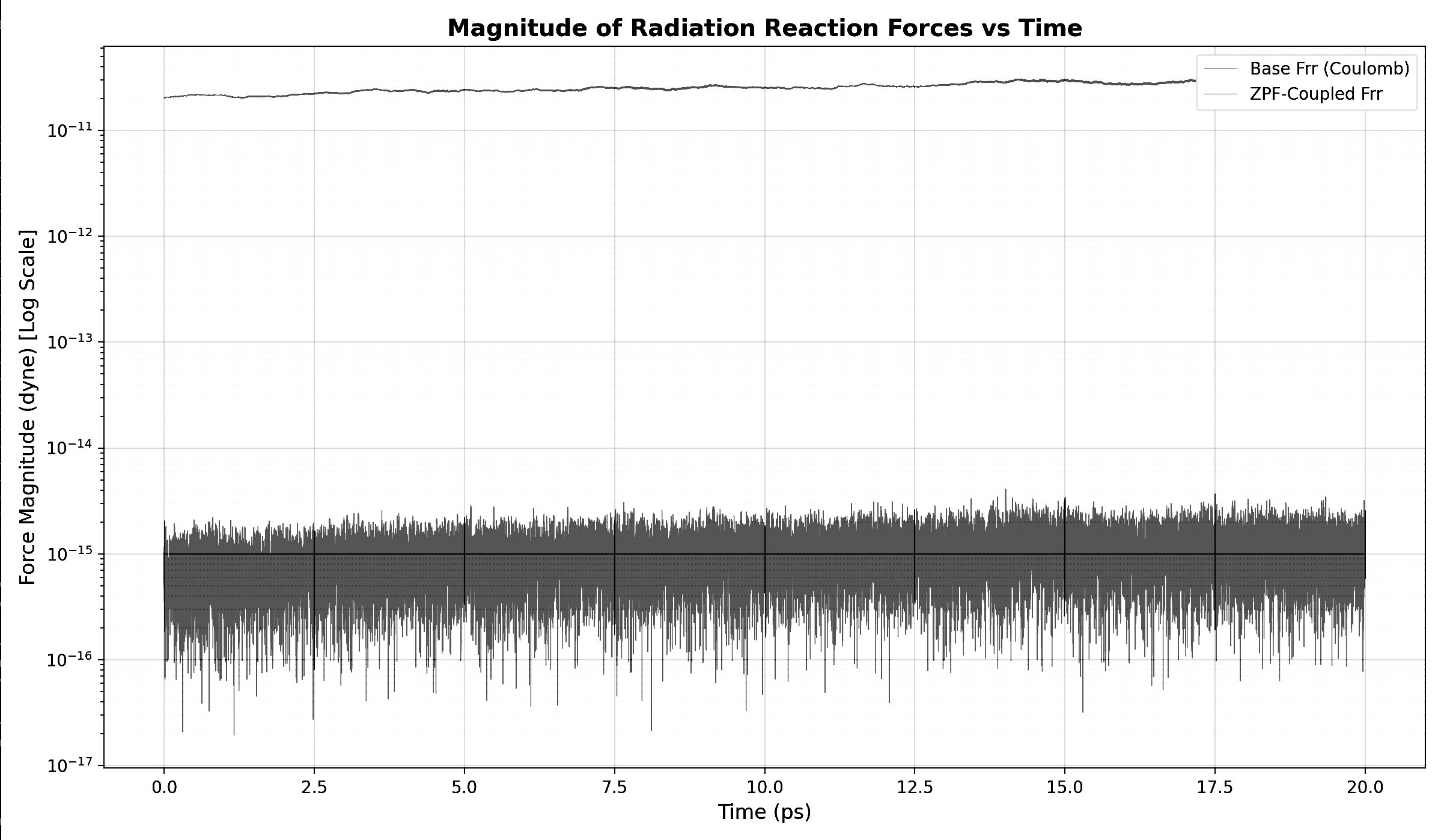}

\textbf{Fig. 1:} The top curve is the magnitude of the radiation reaction
due only to the Coulombic force. \ The bottom curve is the magnitude of the
ZP Lorentz force part of the radiation reaction.
\end{center}

\subsection{Tests on radiation reaction and dipole approximation using the
SHO}

The 3D SHO model, with a resonant frequency of $\omega _{0}=4.5\times 10^{16}
$s$^{-1}$, was used to further test the effects of the stochastic zero-point
Lorentz force.  \ During a 0.5 ps run, $r\left( t\right) $ varied from
approximately $3\times 10^{-10}$cm to $1\times 10^{-11}$cm, due to the
fluctuations induced by the stochastic ZP Lorentz force. \ Over this period
of 0.5 ps, the variance $\left\langle r^{2}\right\rangle $ with the ZP
radiation included in $\vec{F}_{\text{rr}}$ minus the case of $\vec{F}_{%
\text{rr}}$ \textbf{without} the ZP radiation included was of the order of $%
1.5\times 10^{-23}$cm$^{2}$, reflecting the small contribution of including
the ZP Lorentz force in $\vec{F}_{\text{rr}}$, as shown in the previous
subsection. \ 

However, what was even more interesting was the following test. \ The same
SHO simulation was run, but now with $\vec{F}_{\text{rr}}$ due only to the
Coulombic force, and now the evaluation of $\left\langle r^{2}\right\rangle $
with and without the dipole approximation being made. \ The difference
between these two cases over the same 0.5 ps time period, was approximately $%
1.5\times 10^{-26}$cm$^{2}$ , about three orders of magnitude smaller than
the previous test of including the ZP Lorentz force in $\vec{F}_{\text{rr}}.$

Thus, according to these tests, both situations of (1) making the dipole
approximation and (2) not including the ZP Lorentz force in $\vec{F}_{\text{%
rr}}$, were each very small effects and made very good approximations. \
However, \#1 of making the dipole approximation had an even smaller effect
than \#2 not including the ZP Lorentz force in $\vec{F}_{\text{rr}}$,
significantly so, by many orders of magnitude, at least for the SHO\
conditions described here.

\section{Overview of software, code details, and simulation examples}

\subsection{Software overview}

The simulation program is written in C++ and compiled using Xcode. \ Program
execution and input variables are controlled entirely through a config.json
(JavaScript Object Notation) file, enabling users to adjust simulation
conditions without recompiling the source code. There are approximately 25
input parameters available to dictate the simulation's behavior. These
include seven parameters governing the adaptive Runge-Kutta integrator, the
initial radial distance of the electron's orbit, toggles to enable or
disable the ZP fields and/or applied EM fields (such as laser pulses), and
the data output frequency for the integration points. \ Data visualization
and statistical analysis are handled via post-processing Python scripts. \
For example, tracking how the electron's orbit evolves over
time---specifically its semimajor axis $a$ and eccentricity $\varepsilon $%
---is a primary focus. \ The scripts construct a histogram representing the
duration the electron spends at each radial distance from the nucleus, which
is then directly compared to the analytical QM probability distribution of
Eq. (\ref{Pqm(r)}). \ As described in the following subsections, Figures
2--4, which depict these various physical quantities, were generated
entirely using this Python post-processing suite.

\subsection{Several simulation outputs}

The following figures illustrate the capabilities and stability of the
numerical architecture developed in this work. While longer integration
times will be employed in future physics-focused studies, the present 25 ps
ensemble runs clearly demonstrate the direction and potential of this work.
\ All presented results strictly leverage the computational advancements
described in Sections 3.0 and 4.0.

Figure 2 illustrates the time evolution of a classical electron orbiting the
nucleus (proton) in the presence of electromagnetic ZP radiation over a 25
ps interval. The electron was initialized in a circular orbit of radius $%
r=0.53\unit{\mathring{A}}$. The top panel displays the orbital radius $%
r\left( t\right) $ as a function of time, while the second panel tracks the
semi-major axis $a\left( t\right) $. These top two plots appear nearly
identical because the eccentricity $\varepsilon \left( t\right) $ (third
panel) remains remarkably small throughout the simulation ($\varepsilon =0$
corresponds to a circular orbit). The fourth panel records the polar angle
of the normal to the orbital plane in radians; this angle remains below
approximately 43 degrees (0.75 rad). Thus, the orbital plane changes only
slowly across this run of nearly 190,000 orbital revolutions.

\begin{center}
\includegraphics[width=0.9\textwidth]{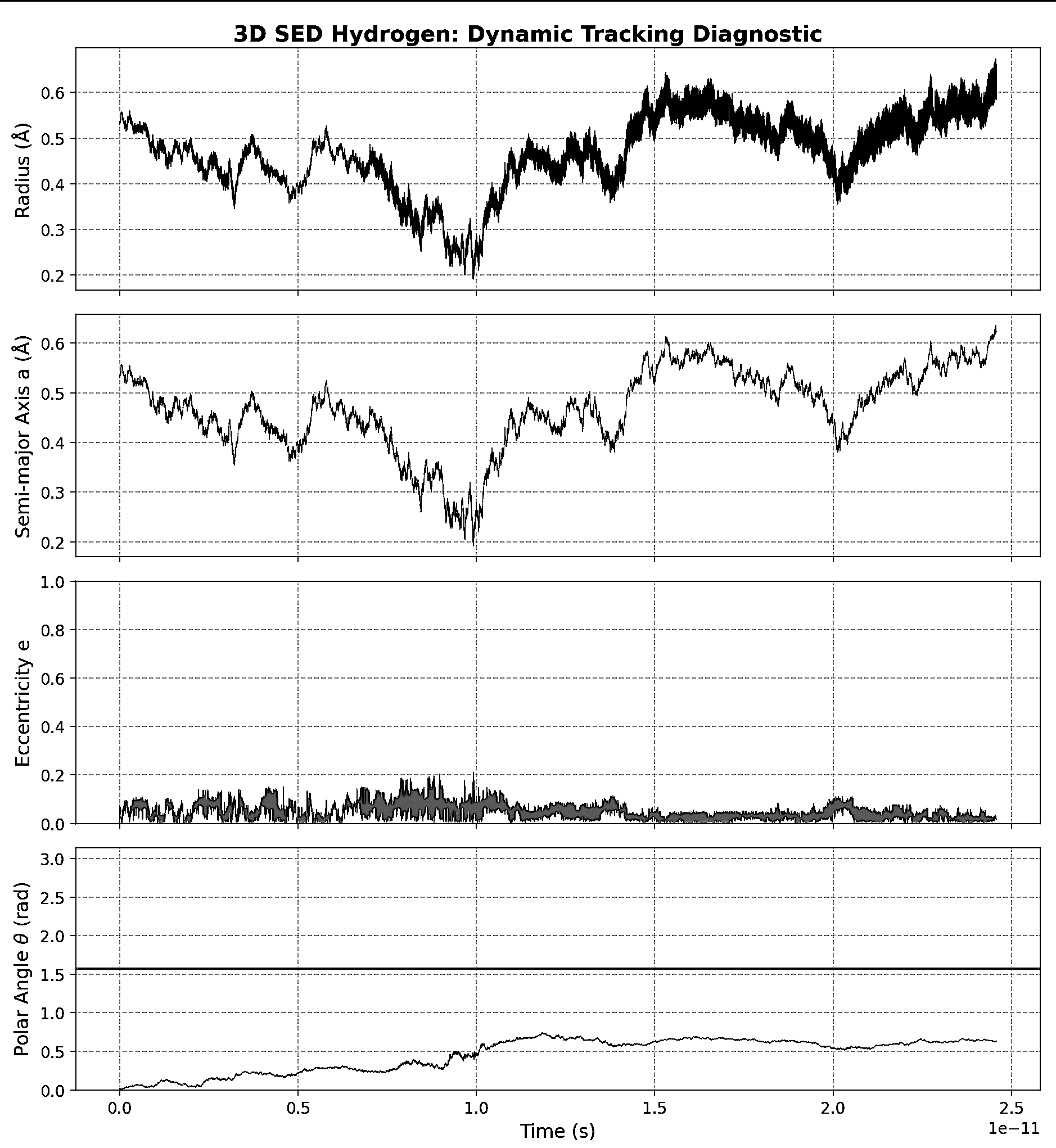}

\textbf{Fig. 2:} Tracking diagnostic of the simulated electron's trajectory.
The top plot shows the radial distance $r$, the distance from the radius of
the nucleus as a function of time $t$. \ The second, third, and fourth
panels display the semimajor axis $a$, the eccentricity $\varepsilon $, and
the polar angle of the orbital plane's normal vector versus time,
respectively.
\end{center}

Figure 3 takes the data of $r\left( t\right) $ vs. $t$ from 17 independent
25 ps simulation runs to determine the likelihood of finding the electron at
each radial position. \ Obtaining this histogram for a single electron
requires tracking the trajectory line crossings at each radial value
(effectively taking horizontal slices of the top plot in Fig. 2) and
adjusting for the time spent in each bin via $\left\vert dr/dt\right\vert $.
\ Clearly, Fig. 2 helps to illustrate how Fig. 3 arises from a classical
physics diffusion-like process driven by the stochastic interaction with ZP
radiation.

Figure 3 is likely the plot of most interest from our simulator, as it
directly compares the probability density $P\left( r\right) $ of finding the
classical electron at a distance $r$ from the nucleus to the analytical
prediction of QM (solid line), Eq. (\ref{Pqm(r)}). \ Both the net area under
the simulation histogram and the curve of Eq. (\ref{Pqm(r)}) are normalized
to unity. \ Crucially, no atomic ionization\ events occurred during any of
the 17 simulation runs. The histogram shows strong future promise for even
tighter agreement with Eq. (\ref{Pqm(r)}) as integration times are extended
beyond 25 ps, since longer durations will help \textquotedblleft
erase\textquotedblright\ the initial $r=0.53\unit{\mathring{A}}$ memory of
the ensemble, allowing further spatial spread to the left and right of the
present distribution. \ Satisfyingly, the picosecond time domain required
here to obtain meaningful 3D results agrees in order of magnitude with the
time span from our 2003 2D simulations in \cite{ColeWilsonPRLett2003}. \ 

\begin{center}
\includegraphics[width=0.9\textwidth]{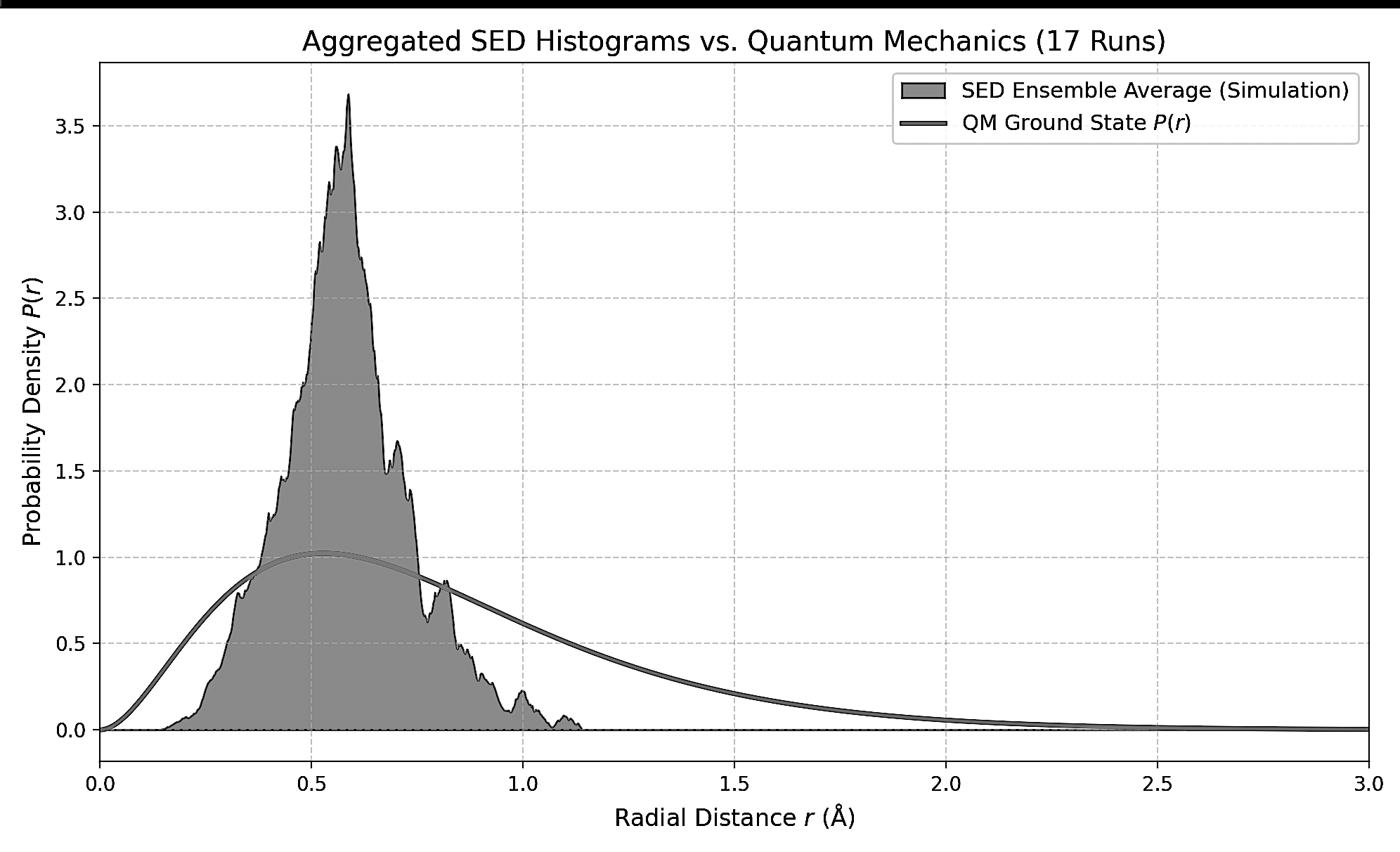}

\textbf{Fig. 3}: Histogram of percent of time an electron spent at each
radius, over the net of 17 electron simulation runs, overlaid with the QM
ground state probability density function.
\end{center}

Further insight is shown in Fig. 4. \ Here the energy $E=\frac{m\vec{v}^{2}}{%
2}-\frac{e^{2}}{r}$ is plotted versus time in the top panel. \ The second
panel displays the magnitude $\left\vert \vec{L}\right\vert $ of the angular
momentum $\vec{L}=\vec{z}\times m\vec{v}$ of the electron in its orbit,
while the third panel shows $L_{z}$, the projection of $\vec{L}$ onto the $z$
axis. \ Since the orbits are slowly changing ellipses, due to the strong
Coulombic attraction, $E\approx -\frac{e^{2}}{2a}$, where $a$ corresponds to
the data tracked in the second panel of Fig. 2.

\begin{center}
\includegraphics[width=0.9\textwidth]{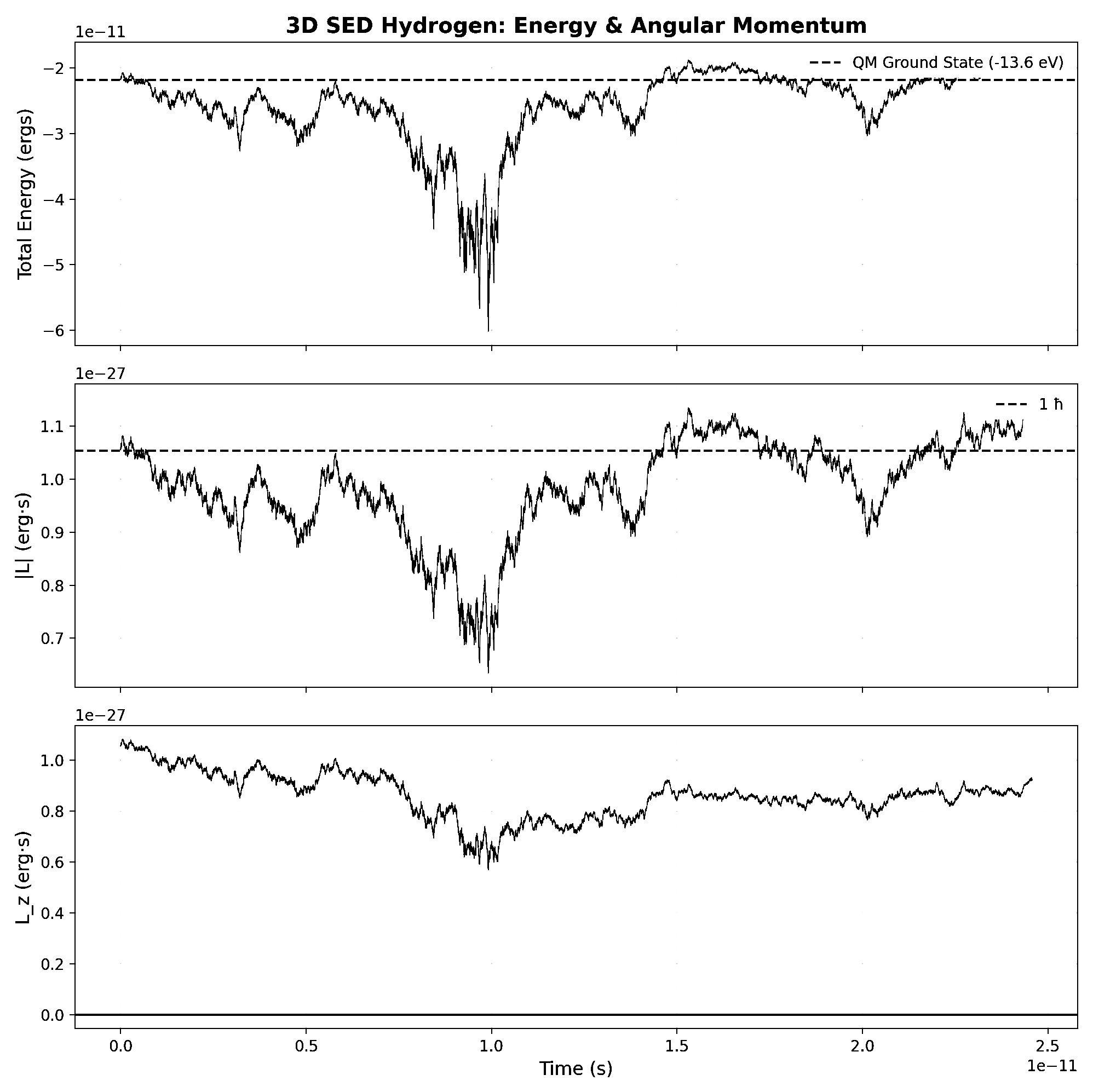}

\textbf{Fig. 4: }Three panels of additional properties of the electron's
\textquotedblleft trajectory\textquotedblright\ ($r$ vs. $t)$ from Fig. 2. \
From top to bottom, respectively, is displayed the energy of the electron,
the magnitude of the angular momentum, and the $z$-component of the angular
momentum, all plotted versus $t$.
\end{center}

\subsection{Software used to investigate artificial electron decays}

The use of this software has already been shown in several ways so far, such
as in several studies in Sect. 5.0, and in the figures produced in this
section. \ While the analytical tests presented in Sect. 5.0 required
specialized modifications to the C++ core during the development phase, all
results and studies detailed in Sect. 6.0 can be reproduced directly using
the publicly released GitHub code. \ Here is an example of such use and
study.

The following convergence study serves as an example of leveraging the
released software to investigate complex physical phenomena. \ Constructing
the ensemble for Fig. 3 provided insight into simulated electron behaviors -
specifically, non-physical orbital collapses. \ In the simulation of Fig. 3,
all electrons were initialized with a circular radius of 0.53$\unit{%
\mathring{A}}$. While most orbits fluctuated and evolved naturally,
occasionally an electron's radius would decrease dramatically and eventually
\textquotedblleft decay\textquotedblright\ into the nucleus. This behavior
is expected in a classical system lacking zero-point (ZP) radiation, but it
contradicts the physical mechanics of SED.

The theoretical expectation was outlined by Boyer in 1975 \cite%
{Boyer1975review1}, who predicted that for small radii and corresponding
high orbital frequencies, the electron resonates with the high-frequency
regime of the ZP spectrum, which scales as $\omega ^{3}$. On average, this
intense radiation resonance should overcome the classical radiation reaction
at that small radius, driving the orbit to increase. Conversely, for larger
radii and lower orbital frequencies, the electron resonates with the
lower-frequency ZP spectrum and should be dominated by radiation reaction,
causing the orbit to decay. Because the process is stochastic, a small orbit
may probabilistically increase or decrease, but the probability of
continuous decay approaches zero since the available ZP energy ($\omega ^{3}$%
) increases rapidly as the orbit shrinks. While this explanation has been
simplified here in terms of circular orbits, simulations to date indicate
that the eccentricity $\varepsilon \left( t\right) $ remains close to zero,
meaning $r(t)$ is approximately the semimajor axis length, the more
appropriate parameter.

To investigate this, a convergence study was conducted (Table 3). As the
number of frequency bins ($N$) representing the ZP field in Eqs. (\ref%
{Sigma-2}) and (\ref{Sigma-3}) is increased, the numerical resolution
improves, and the number of artificial decays decreases. For this study, an
ensemble of 18 electrons was initialized in a tight circular orbit of 0.12 $%
\unit{\mathring{A}}$, representing a highly demanding computational
condition compared to the 0.53 $\unit{\mathring{A}}$ start in Fig. 2. \ A
\textquotedblleft decay\textquotedblright\ was triggered if the electron
collapsed to a threshold of 0.0005$\unit{\mathring{A}}$. The simulation
tracked the ensemble until the first surviving electron reached 1.2 ps. in
simulation time.

\begin{center}
\begin{tabular}{|c|c|c|}
\hline
$N$ (bins) & \textbf{\# Decays} & \textbf{\# Survivors} \\ \hline
1M & 8 & 10 \\ \hline
5M & 7 & 11 \\ \hline
10M & 5 & 13 \\ \hline
50M & 3 & 15 \\ \hline
100M & 3 & 15 \\ \hline
\end{tabular}

\textbf{Table 3: }Number of electron decays vs. number of frequency bins ($N$%
).
\end{center}

As demonstrated, the number of decays systematically decreases as the grid
density increases. However, the execution time scales linearly with the
number of bins. The bulk of the computational overhead involves summing over
the frequencies within the sliding window ($\pm 5\%$ of the central
frequency) for the Lorentz force term at every time step. Consequently,
running a simulation at $N=100M$ takes approximately 100 times longer than $%
N=1M$, making full ensemble studies like Fig. 3 highly impractical at
maximum resolution without access to advanced supercomputing resources.

While Table 3 does not definitively prove that ZP radiation completely
eliminates all classical decays, it is a clear indication that the observed
collapses are numerical artifacts caused by the difficulty of adequately
representing continuous high-frequency spectra in discrete simulations. For
larger starting radii, such as $0.53\unit{\mathring{A}}$, the decay
probability per time unit decreases dramatically, though the probability of
eventual decay inherently increases over long durations when $N$ is made
relatively small. Ultimately, these simulations serve not just to generate
distributions, but as an analytical tool to uncover the subtle physics
governing classical electron trajectories.

\section{Concluding remarks}

Stochastic electrodynamics is a classical theory grounded in a simple but
powerful premise: the necessary equilibrium between classical charged
particles and classical electromagnetic radiation (see Ref. \cite%
{Cole-Alpha-2026}). This dynamic equilibrium necessitates the existence of
classical electromagnetic ZP radiation---a Lorentz-invariant field at
temperature $T=0$ \cite{Marshall1965a}, \cite{Boyer1969Black1} that results
in no heat flow during reversible, isothermal thermodynamic operations \cite%
{Cole19901st}, \cite{Cole-3-charges-2025}. Fundamentally, this ZP radiation
must establish a stable equilibrium for charged particles in an open
environment without containing walls.

Investigating this equilibrium, however, requires advancing beyond the more
conventional, but already nontrivial, treatment of nonlinear oscillator
binding potentials \cite{Boyer1976}, \cite{blanco1}, \cite{blanco2} to focus
strictly on truly realistic atomic systems, such as the classical hydrogen
atom \cite{Boyer1989-Scaling}. This is precisely where the numerical and
computational approaches discussed in this article become essential. Unlike
QM and QED, addressing the physics of this simplest atomic system requires
complex simulation capability. While this may initially seem
counter-intuitive, it mirrors a well-established pattern in physics: when
one foundational theory serves to derive another macroscopic or formalized
theory---much like statistical mechanics deriving thermodynamics, or the
Langevin microscopic equation deriving macroscopic diffusion---the
underlying framework invariably involves significantly greater mathematical
complexity. This is the exact relationship between the hydrogen atom in SED
versus QM. Despite its conceptual elegance, the classical hydrogen atom in
SED is fundamentally intractable without the use of advanced numerical
computation.

This article has outlined the specific mathematical and computational
architecture required to bridge that gap. We have described the symmetrical
consideration of the pseudo-wave formulation, the large-scale indexing
capability of logarithmic frequency gridding, and the handling of resonant
behaviors between radiation and charged particle motion. These
domain-specific advancements operate in tandem with the more recognizable
advanced numerical integration methods required for solving highly nonlinear
differential equations. \ All of these concepts are described in some detail
here. \ They will likely be necessary not just for hydrogen, but also for
more complicated systems such as helium and lithium, where multi-electron
atomic systems are involved. \ Other important considerations involve
externally applied EM fields such as from lasers or strong electric and
magnetic field sources. \ Scaling this methodology to these more complex
systems will likely need to build upon the work described here.

\textbf{Data availability}

The implementation of our method is available on GitHub
[https://github.com/dccole-collab/Hydrogen-3D-Simulator].

\pagebreak

\textbf{Declaration of Generative AI in scientific writing:} During the
preparation of this work, the author used the program Gemini (Google AI) for
aid in computer code generation. After using this AI program, the author
reviewed and edited the work as needed and takes full responsibility for the
content of this research.

\end{document}